\pdfoutput=1
\documentclass[twocolumn,english,amsmath,amssymb,aps,prb,reprint,floatfix,superscriptaddress,showpacs,booktabs]{revtex4}
\usepackage[latin9]{inputenc}
\usepackage{babel}
\usepackage{booktabs}
\usepackage{amsmath}
\usepackage{amssymb}
\usepackage{graphicx}
\usepackage{esint}
\usepackage{subscript}
\usepackage[unicode=true,
 bookmarks=true,bookmarksnumbered=false,bookmarksopen=false,
 breaklinks=false,pdfborder={0 0 1},backref=false,colorlinks=false]
 {hyperref}
\hypersetup{pdftitle={U in pnictides},
 pdfauthor={A. van Roekeghem et al},colorlinks=true,citecolor=blue,breaklinks=true}

\makeatletter

\providecommand{\tabularnewline}{\\}

\@ifundefined{textcolor}{}
{%
 \definecolor{BLACK}{gray}{0}
 \definecolor{WHITE}{gray}{1}
 \definecolor{RED}{rgb}{1,0,0}
 \definecolor{GREEN}{rgb}{0,1,0}
 \definecolor{BLUE}{rgb}{0,0,1}
 \definecolor{CYAN}{cmyk}{1,0,0,0}
 \definecolor{MAGENTA}{cmyk}{0,1,0,0}
 \definecolor{YELLOW}{cmyk}{0,0,1,0}
}

\pdfoutput=1

\usepackage{babel}

\providecommand{\tabularnewline}{\\}

\@ifundefined{textcolor}{}{%
 \definecolor{BLACK}{gray}{0}
 \definecolor{WHITE}{gray}{1}
 \definecolor{RED}{rgb}{1,0,0}
 \definecolor{GREEN}{rgb}{0,1,0}
 \definecolor{BLUE}{rgb}{0,0,1}
 \definecolor{CYAN}{cmyk}{1,0,0,0}
 \definecolor{MAGENTA}{cmyk}{0,1,0,0}
 \definecolor{YELLOW}{cmyk}{0,0,1,0}
}

\makeatother

\begin{document}

\title{Hubbard interactions in iron-based pnictides and chalcogenides: Slater
parametrization, screening channels and frequency dependence}

\author{Ambroise van Roekeghem}

\affiliation{Centre de Physique Théorique, Ecole Polytechnique, CNRS UMR 7644,
Université Paris-Saclay, 91128 Palaiseau, France}

\affiliation{Beijing National Laboratory for Condensed Matter Physics, and Institute
of Physics, Chinese Academy of Sciences, Beijing 100190, China }

\email{\\*\mbox{ambroise.van-roekeghem@polytechnique.edu}}

\selectlanguage{english}%

\author{Lo\"ig Vaugier}

\affiliation{Centre de Physique Théorique, Ecole Polytechnique, CNRS UMR 7644,
Université Paris-Saclay, 91128 Palaiseau, France}

\author{Hong Jiang}

\affiliation{College of Chemistry and Molecular Engineering, Peking University,
100871 Beijing, China}

\author{Silke Biermann}

\affiliation{Centre de Physique Théorique, Ecole Polytechnique, CNRS UMR 7644,
Université Paris-Saclay, 91128 Palaiseau, France}

\affiliation{Collège de France, 11 place Marcelin Berthelot, 75005 Paris, France}

\date{today}
\begin{abstract}
We calculate the strength of the frequency-dependent on-site electronic
interactions in the iron pnictides LaFeAsO, BaFe\textsubscript{2}As\textsubscript{2 },
BaRu\textsubscript{2}As\textsubscript{2 }, and LiFeAs and the chalcogenide
FeSe from first-principles within the constrained random phase approximation.
We discuss the accuracy of an atomic-like parametrization of the two-index
density-density interaction matrices based on the calculation of an
optimal set of three independent Slater integrals, assuming that the
angular part of the Fe-\textit{d} localized orbitals can be described
within spherical harmonics as for isolated Fe atoms. We show that
its quality depends on the ligand-metal bonding character rather than
on the dimensionality of the lattice: it is excellent for ionic-like
Fe-Se (FeSe) chalcogenides and a more severe approximation for more
covalent Fe-As (LaFeAsO, BaFe\textsubscript{2}As\textsubscript{2})
pnictides. We furthermore analyze the relative importance of different
screening channels, with similar conclusions for the different pnictides
but a somewhat different picture for the benchmark oxide SrVO\textsubscript{3}:
the ligand channel does not appear to be dominant in the pnictides,
while oxygen screening is the most important process in the oxide.
Finally, we analyze the frequency dependence of the interaction. In
contrast to simple oxides, in iron pnictides its functional form cannot
be simply modeled by a single plasmon, and the actual density of modes
enters the construction of an effective Hamiltonian determining the
low-energy properties. 
\end{abstract}

\pacs{71.27.+a,71.10.Fd,71.15.Ap,71.45.Gm}

\maketitle

\section{INTRODUCTION}

The discovery of unconventional superconductivity with critical temperatures
up to 55 K in iron-based pnictides and chalcogenides has raised tremendous
interest in the electronic properties of the various classes of these
materials (for recent reviews, see \onlinecite{Hirschfeld-CRAS,
Albenque-CRAS, Ambroise-CRAS}). The first prototypes were LaFeAsO
\cite{Kamihara-2008,Ren-2008,CeOFeAs-chen,LaOFeAs-Wen,SmOFeAs-Chen},
representing the ``1111'' family and the ``122'' compound BaFe$_{2}$As$_{2}$
\cite{BaFeAs-rotter,BaFeSe-sefat,BaFeRuAs-sharma}. Soon, however,
also the ``111'' stoichiometry (such as in LiFeAs \cite{Tapp-LiFeAs,Wang-LiFeAs,Pitcher-LiFeAs}),
and the ``11'' chalcogenides FeSe and FeTe \cite{FeSe-Hsu,FeSe-underpressure-2009,FeSeTe-yeh})
got into the spotlight.

While an obvious difference to the high-temperature superconducting
cuprates is the metallic nature of the undoped iron-pnictides, this
observation does not \textit{a priori} justify a weak-coupling approach,
and several recent works \cite{Medici-selective-Mott,Imada-correlation-pnictides}
suggest a possibly much closer connection to the cuprates than anticipated.
Quite generically, most iron pnictide compounds exhibit magnetically
ordered phases in close proximity to the superconducting ones, and
the relation between the former and the latter remains an open question.
Early models based on a strong coupling picture have met some success
in describing the nature of magnetic ordering, when invoking a biquadratic
exchange term \cite{Yaresko-magnetic-FS}.

On the other hand, early on, a puzzle concerning the value of the
measured magnetic moments was pointed out, namely a magnetic moment
much smaller than the one expected from a high spin configuration
for the Fe 3\textit{d} shell, or from density functional theory (DFT)
calculations \cite{pnictides-mazin,FeSe-alaska}. Within a purely
local picture, the large Hund's coupling, dominating over crystal
and ligand field splittings, would be expected to induce a high-spin
configuration. LaFeAsO, for example, exhibits an antiferromagnetic
local moment between $0.3-0.6\mu_{B}$ below $T\sim130$ K \cite{ishida-2009,LaOFeAs-qureshi-2010},
much smaller than the magnetic moment calculated within DFT ($2\mu_{B}$).
This anomaly was interpreted as a solvation effect due to the extremely
large polarisability of the arsenic ligands \cite{LaOFeAs-Sawatzky},
which leads to a strong reduction of the Hubbard interaction on the
Fe \textit{d} manifold. In ``122'' and in ``11'' families, on
the other hand, larger magnetic moments were determined: around $0.9\mu_{B}$
for BaFe$_{2}$As$_{2}$ \cite{BaFeAs-huang} and $2.2\mu_{B}$ for
FeTe \cite{FeTe-mu-li}. Recent calculations of two-particle correlation
functions within local density approximation (LDA) combined with dynamical
mean field theory (DMFT) have rather suggested dynamic quantum fluctuations
to be the origin of these puzzles, inducing a ``dichotomy'' between
large local but small ordered magnetic moments as measured within
neutron experiments \cite{LaOFeAs-hansmann-2010,LaOFeAs-hansmann-2012}.
Yet another example where the timescale of the experimental probe
is decisive for the outcome of a measurement was recently also analyzed
in Ref. \onlinecite{Hansmann_SciRep}.

Determining the strength of electronic Coulomb correlations appears
therefore an important issue not only for understanding their role
for electronic, magnetic and transport properties but even for establishing
the framework and language in which those are best described.

An additional difficulty arises from a particularity of the iron pnictides,
which -- due to their closeness to a filling of 6 electrons in 5 orbitals
and substantial Hund's rule coupling -- exhibit an extreme sensitivity
with respect to small changes in parameters, and compounds with moderate
electronic correlations can be tuned into rather strongly correlated
ones by modest changes in interaction strength, in particular Hund's
coupling, doping, or composition.

To address such questions, it is mandatory to construct realistic
Hamiltonians to which many-body tools can be applied. The discovery
of the iron pnictides has thus opened an important new testbed for
\textit{ab initio} techniques for correlated materials. Even when
restricting to dynamical mean field theory-based calculations \cite{RevDMFT_AG,kotliar-review-DMFT,LDA+DMFT-anisimov-1997,LDA+DMFT-licht},
the early literature is abundant \cite{LaOFeAs-anisimov-2009,LaOFeAs-shorikov,LaOFeAs-anisimov-2008,cRPA-DMFT-LaOFeAs-markus,LaOFeAs-haule-2008,LaOFeAs-kotliar-2009,liebsch-FeSe,cRPA-DMFT-FeSe-markus,sc-LDA+DMFT-haule}.
Many of the early works were based on one-particle Hamiltonians derived
from density functional theory, supplemented by many-body interaction
terms which were taken as adjustable parameters. It became soon clear,
however, that this is an insufficient strategy for a truly materials-specific
description of iron-pnictides. The determination -- from first principles
-- of the effective local Coulomb interactions is therefore an important
intermediate goal for quantitative theories of iron pnictides.

A most promising route is the constrained random phase approximation
(cRPA) \cite{cRPA-ferdi-2004} -- an approach for deriving from first-principles
the interacting Hamiltonian within a target subspace that is appropriate
for describing the low-energy many-body properties (``downfolding'').
Refs. {[}\onlinecite{cRPA-LaOFeAs-miyake, cRPA-DMFT-LaOFeAs-markus,
cRPA-pnictides-takashi, udyneff-michele, cRPA-LaOFeAs-nakamura, loig-tesis,
Ambroise-Ba2Ti2Fe2As4O, Ambroise-BaCo2As2, Ambroise-CaFe2As2, BaCo2As2-Nan,
Razzoli-pnictides-correlations, cRPA-DMFT-FeSe-markus, udyn-werner}{]}
demonstrated the usefulness of the cRPA for iron pnictides. For LaFeAsO,
Ref. {[}\onlinecite{cRPA-DMFT-LaOFeAs-markus}{]} considered a Hubbard
Hamiltonian (dubbed ``\textit{d-dp}'') that incorporated both, Fe
3\textit{d} and ligand As and O \textit{p} states as degrees of freedom,
but with a Coulomb energy cost on Fe 3\textit{d} orbitals only. The
effective interactions for this specific low-energy model were calculated
within the cRPA \cite{cRPA-ferdi-2004}, and the many-body Hamiltonian
was solved within LDA+DMFT. Within this scheme, LaFeAsO was described
as a metal with moderate strength of the electronic correlations \cite{cRPA-DMFT-LaOFeAs-markus},
whereas the largest effects were found for $\alpha$-FeSe \cite{cRPA-DMFT-FeSe-markus}
in agreement with recent photoemission experiments \cite{FeSe-PES-Yoshida,FeSe-tamai}.
In addition, Hund's rule coupling, $J$, appeared to play a fundamental
role in the description of the low-energy properties of FeSe, as also
noted by Haule and co-workers \cite{LaOFeAs-kotliar-2009} for characterizing
the coherence-incoherence crossover temperature in LaFeAsO.

Beyond giving first principles estimates for the effective Hubbard
and Hund's interactions for specific materials calculations, the cRPA
enables systematic studies of trends along the series. In this way,
the larger values of $U$ and $J$ in the chalcogenides as compared
to the pnictides were rationalized \cite{cRPA-pnictides-takashi},
based on their electronic structure.

The interpretation of the effective local Hubbard interaction as \textit{partially
screened} interaction, that underlies the cRPA strategy has an interesting
further consequence: since screening is a dynamical process, the partially
screened interactions $U$ and $J$ are also dynamical, that is, frequency-dependent
quantities \cite{Biermann_JPCM_review}. The impact of this energy-dependence
on the low-energy properties and the coupling between electronic and
plasmonic excitations in many-body calculations has been studied recently
in prototypical models and the benchmark oxide SrVO$_{3}$ \cite{udyn-michele,Ambroise-SrVO3,Silke_DD_polarons}
as well as in several transition metal pnictides \cite{udyn-werner,Ambroise-BaCo2As2,Ambroise-CaFe2As2}.
The inclusion of energy-dependent Hubbard interactions within an extended
version of LDA+DMFT \cite{udyn-michele,udyn-werner} leads to a reduction
of the quasi-particle weight at the Fermi energy, compared to standard
many-body techniques, e.g. LDA+DMFT. The spectral weight is shifted
to additional satellites at larger energies, in good agreement with
photoemission experiments. A systematic procedure for constructing
low-energy Hamiltonians that incorporate both, an interacting Hamiltonian
downfolded into a low-energy subspace, and the renormalization of
the one-particle part of the Hamiltonian due to electron-plasmon excitations,
was introduced in Ref. {[}\onlinecite{udyneff-michele}{]}. It consists
in estimating from $U(\omega)$, a plasmon coefficient that reduces
the kinetic energy. The values of such coefficient are between 0.59
and 0.63 for LaFeAsO, FeSe and BaFe$_{2}$As$_{2}$ \cite{udyneff-michele}.

In this paper, the strength of the frequency-dependent on-site electronic
interactions in the iron-based pnictide (LaFeAsO, BaFe\textsubscript{2}As\textsubscript{2 }and
BaRu\textsubscript{2}As\textsubscript{2 }, LiFeAs) and chalcogenide
(FeSe) families is calculated from first-principles within the cRPA
\cite{cRPA-ferdi-2004}. We use the recent implementation of Ref.
{[}\onlinecite{TMO-vaugier}{]}, based on the electronic structure
code Wien2k \cite{blaha_wien2k}. Both Fe \textit{d} and ligand As
(and O in oxypnictides) \textit{p} degrees of freedom are considered
in the construction of the resulting parameter-free low-energy ``\textit{dp-dp}''
Hamiltonian. An effective ``\textit{d-dp}'' Hamiltonian where only
the occupation on Fe-\textit{d} orbitals is affected by the Coulomb
repulsion can then be constructed via the recently proposed ``shell-folding''
procedure \cite{Shell-folding} that we extend here to the frequency-dependent
case. The results of this scheme for the static value of the interaction
parameter are similar to calculations in a \textit{d-dp} model in
which only the transitions from and to the bands with a majority of
\textit{d}-orbital character are cut within constrained-RPA and $U_{dp}$
is neglected \cite{loig-tesis}. On the other hand, the infinite frequency
value is reduced by about 30\%. In agreement with the literature,
we find in Section \ref{sec:Hubbard interactions and Slater parametrization}
that the effective Coulomb interactions for Fe-3\textit{d} shells,
are larger in ``11'' chalcogenides than in ``122'' and ``1111''
pnictides, while the ``111'' are an intermediate case.

The accuracy of an atomic-like parametrization of the two-index density-density
interaction matrices within the ``\textit{dp-dp}'' low-energy Hamiltonian
is discussed in Section \ref{sec:Hubbard interactions and Slater parametrization}.
It is based on the calculation of an optimal set of three independent
Slater integrals, assuming that the angular part of the Fe-\textit{d}
localized orbitals can be described within spherical harmonics as
for isolated Fe atoms (see Section \ref{sec:Method} for an introduction).
We find that the accuracy of this parametrisation depends on the ligand-metal
bonding character rather than on the dimensionality of the lattice:
it is excellent for ionic-like Fe-Se (FeSe) chalcogenides and less
appropriate for more covalent Fe-As (LaFeAsO, BaFe\textsubscript{2}As\textsubscript{2})
pnictides. This illustrates the differences in the sphericity of the
Fe-3\textit{d} Wannier orbitals and in the anisotropy of the screening.

In Section \ref{sec:Screening channels}, we investigate the relative
importance of screening channels which reduce the on-site bare interaction
to the fully screened one. We show that the screening channels are
analogously structured in the pnictide and chalcogenide families,
while this structure is very different in a benchmark oxide, namely
SrVO\textsubscript{3}. The ligand channel does not appear to be responsible
for the dominant screening mechanism in iron pnictides.

Finally, we analyze the frequency dependence of the interaction and
its relation with the values of the free-electron plasmon frequencies
in Section \ref{sec:Frequency dependence}. In contrast to simple
oxides, in iron pnictides its functional form cannot be simply modeled
with a single plasmon, and the actual density of modes enters the
construction of the effective Hamiltonian determining the low-energy
properties, through a renormalisation of the quasiparticle dispersions.

\section{Method}

\label{sec:Method}

\subsection{General framework}

The cRPA \cite{cRPA-ferdi-2004} is a first principles tool to construct
low-energy Hamiltonians with specific ``target'' degrees of freedom.
The main idea consists in identifying Hubbard and Hund's interactions
with matrix elements of a partially screened interaction $W^{r}$
within a set of localized Wannier(-like) orbitals $\{|\phi_{m}\rangle\}$,
with $m$ an orbital quantum number. The partially screened interaction
$W^{r}$ corresponds to the bare Coulomb interaction within the target
low-energy subspace for which explicit many-body calculations are
carried out. Within the full Hilbert space $W^{r}$ is a partially
screened interaction, screened by higher energy degrees of freedom
not included in the target space. The most straightforward option
is to choose as the target space the space spanned by the functions
$\{|\phi_{m}\rangle\}$ but other options are possible and will be
exploited below.

The partially screened interaction $W^{r}$ is calculated by constraining
the polarization such that the screening processes involving the target
states included in the low-energy Hamiltonian are not double counted
in a further many-body calculation. The random phase approximation
gives an explicit expression for the polarization, $P$, in terms
of transitions between occupied and empty states. Within this assumption,
it is possible to calculate the constrained polarization $P^{r}=P-P^{sub}$,
where $P^{sub}$ is the polarization within the target low-energy
subspace only \cite{TMO-vaugier}. It is energy-dependent since screening
is a dynamical process. $W^{r}$ is defined as the interaction screened
by $P^{r}$: 
\begin{eqnarray}
W_{r} & = & \left[1-vP_{r}\right]^{-1}v
\end{eqnarray}

The additional screening taking place in the low-energy subspace then
allows us to recover the fully screened interaction $W$: 
\begin{eqnarray}
W & = & \left[1-vP\right]^{-1}v=\left[1-vP^{r}-vP^{sub}\right]v\nonumber \\
 & = & \frac{v/\left[1-vP^{r}\right]}{1-\left[v/\left[1-P^{r}v\right]\right]P^{sub}}\nonumber \\
 & = & \left[1-W^{r}P^{sub}\right]^{-1}W^{r}
\end{eqnarray}

\subsection{Interaction matrices}

The value of the partially screened interaction $W^{r}$ between local
orbitals is expressed in terms of the four-index interaction matrix
$U_{m_{1}m_{2}m_{3}m_{4}}^{(\mathcal{S})}$: 
\begin{multline}
U_{m_{1}m_{2}m_{3}m_{4}}^{(\mathcal{S})}(\omega)\equiv\langle\phi_{m_{1}}\phi_{m_{2}}|W^{r}(\omega)|\phi_{m_{3}}\phi_{m_{4}}\rangle\\
=\iint d^{3}rd^{3}r'\phi_{m_{1}}^{*}(r)\phi_{m_{3}}(r)W^{r}(r,r';\omega)\phi_{m_{2}}^{*}(r')\phi_{m_{4}}(r')\label{Wcrpa3}
\end{multline}
 where the superscript $\mathcal{S}$ is added for specifying the
angular symmetry of the localized orbitals considered.

Most matrix elements are of the order of 0.1 eV or less, except for
two-index reduced interaction matrices, $U_{mm'}^{\sigma\sigma}|_{\textrm{cRPA}}$,
$U_{mm'}^{\sigma\bar{\sigma}}|_{\textrm{cRPA}}$ and $J_{mm'}^{\textrm{cubic}}$
which can be extracted from the calculation. Cubic angular harmonics
are considered in our case as an approximation to the crystal field
in the iron-based pnictides and chalcogenides: 
\begin{align}
U_{mm'}^{\sigma\bar{\sigma}}|_{\textrm{cRPA}} & \equiv U_{mm'mm'}^{\textrm{cubic}}=\langle\phi_{m}\phi_{m'}|W^{r}(0)|\phi_{m}\phi_{m'}\rangle\label{Umm1}\\
U_{mm'}^{\sigma\sigma}|_{\textrm{cRPA}} & \equiv U_{mm'mm'}^{\textrm{cubic}}-U_{mm'm'm}^{\textrm{cubic}}\label{Umm2}
\end{align}
 where $m$ runs over the \textit{d} orbital subspace and $\sigma$
refers to the spin degree of freedom.

As two atoms of Fe are found in the conventional unit-cell of the
iron-based pnictides and chalcogenides, one has access to the nearest-neighbor
interaction between Fe-3\textit{d} orbitals within equation \ref{Wcrpa3}
(see Ref. {[}\onlinecite{TMO-vaugier}{]} for a more general expression
of the non-local interactions). One can also calculate the interaction
between Fe-3\textit{d} and As-4\textit{p} orbitals in the same way.

\subsection{Slater parametrization}

\label{sub:Slater parametrization}

Replacing the four-index interaction matrix $U_{m_{1}m_{2}m_{3}m_{4}}$
by a small subset of fitting parameters is usually done in the literature
of many-body calculations, e.g. LDA+U or LDA+DMFT, in order to avoid
double-counting issues. However, the errors induced by considering
such an approximated interacting Hamiltonian in many-body calculations
have not been investigated yet%
\footnote{We note that an orbital-dependent double-counting in the Around Mean
Field spirit has been proposed \cite{Ambroise-BaCo2As2} and provided
good agreement with the experiment for a compound with two correlated
shells \cite{Ambroise-Ba2Ti2Fe2As4O}, including for the position
of non-correlated bands.%
}. For isolated atoms, the development into a finite number of Legendre
polynomials of the matrix elements of the Coulomb potential with spherical
harmonics is exact \cite{judd,Slater_book,Sugano_book}. It involves
only three radial integrals -- or Slater integrals -- for \textit{d}
states, whereas the angular part is determined with well defined Racah-Wigner
coefficients, $\alpha_{k}$: 
\begin{multline}
\alpha_{k}(m_{1},m_{2},m_{3},m_{4})\\
=\frac{4\pi}{2k+1}\sum_{q=-k}^{k}\langle Y_{lm_{1}}|Y_{kq}Y_{lm_{3}}\rangle\langle Y_{lm_{2}}Y_{kq}|Y_{lm_{4}}\rangle\label{Racah}
\end{multline}
 where $Y_{lm}$ are spherical harmonics and $\langle Y_{l_{1}m_{1}}|Y_{l_{2}m_{2}}Y_{l_{3}m_{3}}\rangle$
refer to the Gaunt coefficients. It is the sphericity of the isolated
atom -- and of the spherical harmonics used -- that sets the finite
number of the Slater integrals to $l+1$ where $l$ is the orbital
quantum number.

Assuming that i) the localized Wannier orbitals, $\{|\phi_{m,-2\leq m\leq2}\rangle\}$,
to which the Hamiltonian is downfolded at low-energy, still retain
the sphericity of the isolated atom although they are embedded in
the solid, and ii) screening does not induce strong orbital anisotropy,
allows to define Slater integrals for correlated orbitals in materials
as follows \cite{TMO-vaugier}: 
\begin{multline}
F^{k}(\omega)=\mathcal{C}_{l,k}\sum_{m_{1},m_{2},m_{3},m_{4}}(-1)^{m_{1}+m_{4}}U_{m_{1}m_{2}m_{3}m_{4}}^{\textrm{(spheric)}}(\omega)\\
\times\left({\setlength\arraycolsep{1pt}\begin{array}{ccc}
l & k & l\\
-m_{1} & m_{1}-m_{3} & m_{3}
\end{array}}\right)\left({\setlength\arraycolsep{1pt}\begin{array}{ccc}
l & k & l\\
-m_{2} & m_{2}-m_{4} & m_{4}
\end{array}}\right)\label{slaterFU1}
\end{multline}
 where the parentheses correspond to the Wigner 3j-symbols and the
coefficients $\mathcal{C}_{l,k}$ are defined as follows: 
\begin{eqnarray}
\mathcal{C}_{l,k} & = & \frac{2k+1}{(2l+1)^{2}\left({\setlength\arraycolsep{2pt}\begin{array}{ccc}
l & k & l\\
0 & 0 & 0
\end{array}}\right)^{2}}.
\end{eqnarray}
 The superscript ``spheric'' indicates that Wannier orbitals with
spherical angular harmonics are employed. The usual definition of
the Hubbard $U=F^{0}$ and Hund's exchange $J=(F^{2}+F^{4})/14$ follows.
Analogously, we define the bare parameters, $v$ and $J_{\textrm{bare}}$
when considering the Slater integrals that parametrize the bare interaction
matrix elements.

The Slater integrals can be used for calculating the Slater-symmetrized
interaction matrix, $\bar{U}_{m_{1}m_{2}m_{3}m_{4}}^{(\mathcal{S})}$,
with the symmetry $\mathcal{S}$ of the crystal field: 
\begin{multline}
U_{m_{1}m_{2}m_{3}m_{4}}^{(\mathcal{S})}(\omega)=\sum_{m_{1}'m_{2}'m_{3}'m_{4}'}\mathcal{S}_{m_{1}m_{1}'}\mathcal{S}_{m_{2}m_{2}'}\\
\times\bigg\{\sum_{k=0}^{2l}\alpha_{k}(m_{1}',m_{2}',m_{3}',m_{4}')F^{k}(\omega)\bigg\}\mathcal{S}_{m_{3}'m_{3}}^{-1}\mathcal{S}_{m_{4}'m_{4}}^{-1}\label{slaterFU2}
\end{multline}
 Choosing $\mathcal{S}$ as the transformation from spherical to cubic
harmonics leads to the Slater-symmetrized reduced interaction matrices
with cubic symmetry: 
\begin{align}
\bar{U}_{mm'}^{\sigma\bar{\sigma}}|_{\textrm{Slater}} & \equiv\bar{U}_{mm'mm'}^{\textrm{cubic}}\label{Uslater1}\\
\bar{U}_{mm'}^{\sigma\sigma}|_{\textrm{Slater}} & \equiv\bar{U}_{mm'mm'}^{\textrm{cubic}}-\bar{U}_{mm'm'm}^{\textrm{cubic}}.\label{Uslater2}
\end{align}

We stress that within this method three independent Slater integrals
are deduced for \textit{d} Hubbard interaction matrices, see Ref.
{[}\onlinecite{loig-tesis}{]} for details. This allows for an unbiased
check of the commonly used assumption of setting the ratio $F^{4}/F^{2}$
to a fixed value of $0.63$ for 3d orbitals, leaving only two independent
Slater integrals (see e.g. the discussion in \cite{LDA+U-anisimov-1997}).
Relations similar to equation \ref{slaterFU1} were used in Ref. {[}\onlinecite{scGW-kutepov}{]}
for BaFe$_{2}$As$_{2}$, based on a self-consistent GW approximation
for calculating the four-index Hubbard interaction matrix.

\subsection{Frequency dependence}

\label{sub:Frequency dependence}

Because of the frequency dependence of the constrained polarization,
the partially screened interaction $W^{r}$ is also frequency dependent.
Consequently, the $U$ matrix and the Slater integrals parametrizing
it are defined as a function of frequency: $U=U(\omega)$. At infinite
frequency, the interaction $W^{r}(\omega=\infty)$ is equal to the
bare, unscreened Coulomb interaction $v$. The largest variation is
observed on the monopole part $F_{0}$, which can be reduced by one
order of magnitude at zero frequency compared to the unscreened value,
while the multipole terms are proportionally less impacted \cite{U-xray-sawatzky-1977,NiO-sawatzky,U-xray-sawatzky-1984}.

Naively, one might think that the high-frequency tail should have
little influence on the low-energy spectral properties, since typical
plasmon frequencies are usually the largest energy scale in the problem.
This is however not true, due to the mechanism alluded to above: the
frequency-dependence can be understood as resulting from a coupling
of the electrons to bosonic screening degrees of freedom, and the
resulting eigenstates of the coupled fermion-boson problem can be
understood as ``electronic polarons'', electrons dressed by their
bosonic screening cloud. These entities have larger effective masses
and thus renormalised dispersions. An explicit construction of an
effective low-energy Hamiltonian incorporating these renormalisation
has been derived in Ref. \onlinecite{udyneff-michele}. The idea
is to introduce a bosonic renormalization factor Z\textsubscript{B}
accounting for the screening modes \cite{udyneff-michele}. In the
general form for the dynamical interaction $\frac{1}{2}\left(V\delta(\tau)+U_{ret}(\tau)\right)n(\tau)n(\tau')$,
the screening is contained in U\textsubscript{ret} while V corresponds
to the bare interaction. Introducing the screening modes of energy
$\omega$ and coupling strength $\lambda(\omega)=\sqrt{-\Im(U_{ret}(\omega))/\pi}$
allows us to parametrize U\textsubscript{ret} as: 
\begin{equation}
U_{ret}(\tau)=-\int_{0}^{\infty}d\omega\lambda^{2}(\omega)\cosh\left[(\tau-\frac{\beta}{2})\omega\right]/\sinh\left[\frac{\beta\omega}{2}\right]
\end{equation}
 and we can then write the Hamiltonian as a Hubbard-Holstein model:
\begin{eqnarray}
H & = & -\sum_{ij\sigma}t_{ij}d_{i\sigma}^{\dagger}d_{j\sigma}+V\sum_{i}d_{i\uparrow}^{\dagger}d_{i\uparrow}d_{i\downarrow}^{\dagger}d_{i\downarrow}+\mu\sum_{i\sigma}d_{i\sigma}^{\dagger}d_{i\sigma}\nonumber \\
 &  & +\int_{0}^{\infty}\omega\sum_{i}b_{i}^{\dagger}(\omega)b_{i}(\omega)d\omega\nonumber \\
 &  & +\int_{0}^{\infty}\lambda(\omega)\sum_{i\sigma}d_{i\sigma}^{\dagger}d_{i\sigma}\left(b_{i}(\omega)+b_{i}^{\dagger}(\omega)\right)d\omega
\end{eqnarray}
 where i,j are the index of the lattice sites, $t_{ij}$ is the hopping
amplitude between sites i and j, $\mu$ the chemical potential of
the system, $d_{i\sigma}^{\dagger}$ ($d_{i\sigma}$) the creation
(annihilation) operator of electrons of spin $\sigma$ on site i and
$b_{i}^{\dagger}(\omega)$ ($b_{i}(\omega)$) the creation (annihilation)
operator of a quantum of energy in the bosonic mode of energy $\omega$.

Applying a generalized Lang-Firsov transformation to the model \cite{Lang_Firsov_transf,Werner-Holstein}
and projecting onto the subspace of zero-boson states (an approximation
valid at low energies) finally provides us with the following Hamiltonian:

\begin{equation}
H_{eff}=-\sum_{ij\sigma}Z_{B}t_{ij}d_{i\sigma}^{\dagger}d_{j\sigma}+U_{0}\sum_{i}d_{i\uparrow}^{\dagger}d_{i\uparrow}d_{i\downarrow}^{\dagger}d_{i\downarrow}
\end{equation}

Where U\textsubscript{0} is the static value of the Coulomb interaction
and Z\textsubscript{B} reflects the density of screening modes $\frac{\Im U_{ret}(\omega)}{\pi\omega^{2}}$:

\begin{equation}
\ln\left(Z_{B}\right)=-\int_{0}^{+\infty}\frac{\Im U_{ret}(\omega)}{\pi\omega^{2}}d\omega
\end{equation}

Physically, it implies that at low energy the spectral function is
further renormalized by Z\textsubscript{B}, and the remaining weight
is transfered to higher energy. Moreover, the hopping amplitude between
non-correlated and correlated states will be reduced by a factor $\sqrt{Z_{B}}$.

\subsection{Shell-folding}

\label{sub:shell-folding}

At infinite frequency screening is suppressed and the instantaneous
intrashell Coulomb interaction in iron pnictides, $U^{dd}(\omega=\infty)$,
is around 20 eV. This is about one order of magnitude bigger than
the static interaction $U^{dd}(\omega=0)$. At the same time, the
intershell \textit{p-d} interaction $U^{dp}(\omega=\infty)$ is of
the order of 6 eV, and it would seem unreasonable to neglect it. Indeed,
the pd-interaction can provide an important screening mechanism, since
adding charge on the $d$-shell can push charge out of the $p$-shell,
thus reducing the electron addition cost. This mechanism is familiar
since the early ideas of Herring \cite{herring} on ``perfect screening''.
As discussed recently \cite{Shell-folding}, in the context of the
cRPA it can be used to construct a ``shell-folding' scheme that allows
to include $pd$-screening even in situations where entanglement between
$d$- and $p$-states makes the standard ``d-dp'' procedure of the
cRPA ill-defined. We briefly review the main idea, since in the later
sections we will give results both using the standard procedure and
the shell-folded scheme. In particular, we will use an extended version
of shell-folding in the frequency-dependent case.

The main idea can be understood by considering the following purely
algebraic manipulation: We start from a \textit{dp} model, where the
interaction part of the Hamiltonian on one site reads: 
\begin{eqnarray}
H_{int} & = & \frac{1}{2}\sum_{\begin{subarray}{c}
(m,\sigma)\neq(m',\sigma')\\
m,m'\in\{d\}
\end{subarray}}U_{m\sigma m'\sigma'}^{dd}n_{m\sigma}n_{m'\sigma'}\nonumber \\
 & + & \frac{1}{2}\sum_{\begin{subarray}{c}
(m,\sigma)\neq(m',\sigma')\\
m,m'\in\{p\}
\end{subarray}}U_{m\sigma m'\sigma'}^{pp}n_{m\sigma}n_{m'\sigma'}\nonumber \\
 & + & \sum_{\sigma,\sigma'}U^{dp}N_{d\sigma}N_{p\sigma'}.
\end{eqnarray}

This expression is strictly equal to: 
\begin{eqnarray}
H_{int} & = & \frac{1}{2}\sum_{\begin{subarray}{c}
(m,\sigma)\neq(m',\sigma')\\
m,m'\in\{d\}
\end{subarray}}\tilde{U}_{m\sigma m'\sigma'}^{dd}n_{m\sigma}n_{m'\sigma'}\nonumber \\
 & + & \frac{1}{2}\sum_{\begin{subarray}{c}
(m,\sigma)\neq(m',\sigma')\\
m,m'\in\{p\}
\end{subarray}}\tilde{U}_{m\sigma m'\sigma'}^{pp}n_{m\sigma}n_{m'\sigma'}\nonumber \\
 & + & U^{dp}\frac{N(N-1)}{2}
\end{eqnarray}

where $N=\sum_{\sigma}(N_{d\sigma}+N_{p\sigma})$ is the total number
of electrons in \textit{p} and \textit{d} orbitals and $\tilde{U}^{dd}=U^{dd}-U^{dp}$,
$\tilde{U}^{pp}=U^{pp}-U^{dp}$. In many compounds $U^{pp}$ is of
the order of $U^{dp}$, so that the $\tilde{U}^{pp}$ term can be
neglected. If, locally, the dominant screening mechanism is driven
by the $dp$-interaction, one may consider the following assumption:
adding charge onto the $d$-shell pushes away charge from the $p$-shell,
such that the total charge on $d$- and $p$-shells is conserved.
$N$ is then a good quantum number, and the above rewriting corresponds
to a reduction of a $dp$-Hamiltonian to an effective $d-dp$- one.
We end up with a Hubbard model where only the \textit{d} subspace
is considered as correlated with a renormalized Coulomb interaction
\begin{equation}
\tilde{U}^{dd}=U^{dd}-U^{dp}\label{eq:shellfolding}
\end{equation}

The same reasoning can be carried out in the presence of frequency-dependent
interactions, and in Section \ref{sec:Frequency dependence} we will
study the frequency-dependence of the resulting shell-folded interaction.
In the following, we will discuss the shell-folded matrices, but simplify
the notation such as to drop the tildes and superscripts. If nothing
else is indicated, $U$ will therefore mean the $dd$-part of the
matrix.

\section{Hubbard interactions and Slater parametrization in pnictides and
chalcogenides}

\label{sec:Hubbard interactions and Slater parametrization}

\subsection{General trends}

\begin{table}
\caption[Lattice parameters and energy windows used for the construction of
Wannier functions]{Lattice parameters used for the iron pnictides and chalcogenides
and energy windows $\mathbb{W}_{dp}$ (in eV) for the \textit{d-dp}
low-energy Hamiltonians. \textit{d} localized orbitals are constructed
out of the Kohn-Sham states included in $\mathbb{W}_{dp}$.{\footnotesize \label{windows}}}

\centering{}{\footnotesize }%
\begin{tabular*}{1\columnwidth}{@{\extracolsep{\fill}}lcccc}
\toprule[0.08em]  & a(Å) & c(Å) & $\text{z}_{As}$ & $\mathbb{W}_{dp}$(eV)\tabularnewline
\midrule[0.05em] FeSe & 3.77 & 5.50 & 0.267 & {[}-6.5,2.4{]}\tabularnewline
LiFeAs & 3.79 & 6.36 & 0.2635 & {[}-6.0,2.8{]}\tabularnewline
BaFe$_{2}$As$_{2}$ & 3.96 & 13.02 & 0.3545 & {[}-6.5,2.7{]}\tabularnewline
LaFeAsO & 4.03 & 8.74 & 0.349 & {[}-5.5,2.5{]}\tabularnewline
BaRu$_{2}$As$_{2}$ & 4.15 & 12.25 & 0.353 & {[}-6.5,3.6{]}\tabularnewline\bottomrule[0.08em] 
\end{tabular*}
\end{table}

We calculate the four-index-Coulomb interaction matrices $U_{m_{1}m_{2}m_{3}m_{4}}$
from first-principles for a \textit{dp-dp} Hamiltonian. Most matrix
elements are of the order of 0.1 eV or less, except for two-index
reduced interaction matrices, $U_{mm'}^{\sigma\sigma}|_{\textrm{cRPA}}$,
$U_{mm'}^{\sigma\bar{\sigma}}|_{\textrm{cRPA}}$ and $J_{mm'}^{\textrm{cubic}}$
which can be extracted from the calculation. We then apply the shell-folding
procedure described in Section \ref{sub:shell-folding}. The values
for the Hubbard $U$ and Hund's exchange $J$ for the effective \textit{d-dp}
low-energy Hamiltonian are reported in Table \ref{pnictide-1}. Here
$U$ is defined as the mean value of the full $U_{mm'}^{\sigma\bar{\sigma}}$
matrix while $J$ is defined such that $U-J$ is the mean value of
$U_{mm'}^{\sigma\sigma}$. For the latter matrix, the average is taken
over the 20 non-diagonal (and thus non-zero) matrix elements. 4x4x3,
5x5x2, 4x4x2 and 4x4x4 meshes were used for the Brillouin zone integration
for FeSe, LaFeAsO, LiFeAs and BaFe$_{2}$As$_{2}$ and BaRu$_{2}$As$_{2}$,
respectively. The localized orbitals for Fe-3\textit{d} and Ru-4\textit{d}
are constructed out of the Kohn-Sham states within the energy window
$\mathcal{W}_{dp}$ (Table \ref{windows}), within the implementation
of Ref. {[}\onlinecite{cRPA-DMFT-LaOFeAs-markus}{]}.

\begin{table*}
\caption[Coulomb interaction parameters for FeSe, LiFeAs, BaFe$_{2}$As$_{2}$,
LaFeAsO and BaRu$_{2}$As$_{2}$]{Hubbard $U_{eff}(\equiv F^{0}-U_{dp})$, Hund's exchange $J(\equiv(F^{2}+F^{4})/14)$
and screened ratio $F^{4}/F^{2}$ for effective (shell-folded) \textit{d-dp}
Hamiltonians. Both static ($\omega=0$) and infinite frequency values
are shown, as well as the unscreened (bare) interaction $v\equiv F^{0}(\omega=+\infty)$.
The mean value of the intershell interaction $U_{dp}(0)$ and $U_{dp}(+\infty)=v_{dp}$
is also reported, along with the intrashell interaction $U_{pp}(0)$
and $U_{pp}(+\infty)=v_{pp}$. Values in parentheses (see also Ref.
{[}\onlinecite{cRPA-pnictides-takashi}{]}) are indicated for comparison
with cRPA calculations using maximally localized Wannier functions
to represent the \textit{d} local orbitals. We also show the value
$U_{cut-d}(\omega=0)\equiv F_{cut-d}^{0}(\omega=0)$ of the interaction
calculated in an ``entangled'' \textit{d-dp} model where only $d\rightarrow d$
transitions are removed\cite{loig-tesis}.\label{pnictide-1} }

\centering{}{\footnotesize }%
\begin{tabular*}{1\textwidth}{@{\extracolsep{\fill}}lcccccccccccccc}
\toprule 
\toprule[0.08em] (eV) & $U_{eff}$ &  & $U_{cut-d}$ & $J$ &  & $F^{4}/F^{2}$ & $U_{dp}$ & $U_{pp}$ & $U_{eff}(+\infty)$ & $v$ & $J(+\infty)$ & $F^{4}/F^{2}(+\infty)$ & $v_{dp}$ & $v_{pp}$\tabularnewline
\midrule 
\midrule[0.05em] FeSe & 3.90 & (4.0{\footnotesize \cite{cRPA-DMFT-FeSe-markus}}) & 3.97 & 0.92 & (0.9{\footnotesize \cite{cRPA-DMFT-FeSe-markus}}) & 0.699 & 2.11 & 4.02 & 14.32 & 20.36 & 1.03 & 0.623 & 6.04 & 10.18\tabularnewline
LiFeAs & 3.06 &  & 3.03 & 0.86 &  & 0.704 & 1.85 & 3.21 & 13.77 & 19.51 & 0.97 & 0.624 & 5.74 & 8.81\tabularnewline
BaFe$_{2}$As$_{2}$ & 2.30 & (2.7{\footnotesize \cite{udyn-werner}}) & 2.53 & 0.81 &  & 0.725 & 1.33 & 2.42 & 13.73 & 19.31 & 0.96 & 0.620 & 5.58 & 8.39\tabularnewline
LaFeAsO & 1.97 & (2.7{\footnotesize \cite{cRPA-DMFT-LaOFeAs-markus}}) & 2.43 & 0.77 & (0.7{\footnotesize \cite{cRPA-DMFT-LaOFeAs-markus}}) & 0.732 & 1.17 & 2.10 & 13.23 & 18.74 & 0.92 & 0.622 & 5.51 & 8.20\tabularnewline
BaRu$_{2}$As$_{2}$ & 1.80 &  & 2.44 & 0.58 &  & 0.804 & 1.40 & 2.46 & 7.78 & 13.13 & 0.72 & 0.669 & 5.35 & 8.30\tabularnewline\bottomrule[0.08em] 
\bottomrule
\end{tabular*}
\end{table*}

FeSe is the material that exhibits the largest Hubbard $U=3.9$ eV
and Hund's exchange $J=0.9$ eV. We obtain similar values within a
direct calculation of a non-shell-folded \textit{d-dp} model where
only transitions from \textit{d} to \textit{d} bands are cut (see
Ref. {[}\onlinecite{loig-tesis}{]}), which was expected since there
is negligible hybridization between Fe-\textit{d} orbitals and Se-\textit{p}
orbitals. The values agree with the ones calculated within an implementation
of cRPA employing maximally localized Wannier orbitals as Fe-3\textit{d}
local orbitals \cite{cRPA-pnictides-takashi,cRPA-DMFT-FeSe-markus}.
These relatively large (as compared to other pnictides) values for
$U$ and $J$ have been used in LDA+DMFT calculations in Ref. \onlinecite{cRPA-DMFT-FeSe-markus}:
they lead to much more pronounced correlation effects than in iron
pnictide compounds, in agreement with experiments \cite{FeSe-PES-Yoshida,FeSe-tamai,Yamasaki-FeSe}.
In particular, -- in contrast to the iron pnictides -- the calculations
for FeSe found a lower energy feature that was identified as a lower
Hubbard band \cite{cRPA-DMFT-FeSe-markus}. This was confirmed by
spectroscopic findings \cite{FeSe-PES-Yoshida,Yamasaki-FeSe}.

The Hubbard $U$ within the \textit{d-dp} low-energy Hamiltonian in
the iron-based pnictides LaFeAsO and BaFe$_{2}$As$_{2}$ is about
1.5 eV smaller than in the chalcogenide FeSe (see Table \ref{pnictide-1}).
The screened ratios of the Slater integrals, $F^{4}/F^{2}$, on the
other hand, deviate more from the empirical atomic value for 3\textit{d}
shells.

In comparison, the values for the Hubbard $U$ and Hund's coupling
$J$ for BaRu$_{2}$As$_{2}$ are lower than for BaFe$_{2}$As$_{2}$.
Since the Ru-4\textit{d} orbitals are more extended than the Fe-3\textit{d}
(as illustrated by the substantially larger bandwidth of the Ru-4\textit{d}
bands, which is almost 2 eV larger than the one of Fe-3\textit{d}),
the kinetic energy of the Ru-4\textit{d} electrons is more important.
As a consequence, correlations in BaRu$_{2}$As$_{2}$ are weak, and
the DFT-LDA band structure without any further renormalisations is
in good agreement with photoemission experiments \cite{BaFeAs-brouet}.

Finally, LiFeAs can be considered as an intermediate case, where the
Coulomb interactions are higher than in the two other materials. This
trend can be linked to the longer Fe-As distance of 2.42 Å in this
compound, compared to 2.40 Å in the others, resulting in more atomic-like
iron Wannier functions.

\subsection{Accuracy of the Slater parametrization}

We now display the interaction matrices for FeSe, LiFeAs, BaFe\textsubscript{2}As\textsubscript{2}
and BaRu\textsubscript{2}As\textsubscript{2} and discuss the accuracy
of the Slater parametrization introduced in Section \ref{sub:Slater parametrization}.

\subsubsection{FeSe}

Within the basis of cubic harmonics (using the ordering $d_{3z^{2}-r^{2}}$,$d_{x^{2}-y^{2}}$,$d_{xy}$,$d_{xz}$,$d_{yz}$)
the effective local Hubbard interaction matrices (in eV) obtained
starting from a cRPA calculation in a \textit{dp-dp} model (see equations
\ref{Umm1} and \ref{Umm2}) and after shell-folding (see Eq. \ref{eq:shellfolding})
with the intershell interaction $U_{dp}=2.11$ eV read:

\[
U_{mm'}^{\sigma\sigma}|_{\textrm{cRPA}}=\left(\begin{array}{ccccc}
0 & 2.56 & 2.53 & 3.44 & 3.44\\
2.56 & 0 & 3.65 & 2.85 & 2.85\\
2.53 & 3.65 & 0 & 2.81 & 2.81\\
3.44 & 2.85 & 2.81 & 0 & 2.87\\
3.44 & 2.85 & 2.81 & 2.87 & 0
\end{array}\right)
\]
 
\[
U_{mm'}^{\sigma\bar{\sigma}}|_{\textrm{cRPA}}=\left(\begin{array}{ccccc}
4.97 & 3.36 & 3.33 & 3.95 & 3.95\\
3.36 & 4.94 & 4.06 & 3.55 & 3.55\\
3.33 & 4.06 & 4.85 & 3.52 & 3.52\\
3.95 & 3.55 & 3.52 & 4.99 & 3.56\\
3.95 & 3.55 & 3.52 & 3.56 & 4.99
\end{array}\right).
\]

There is a small orbital dependence of the intra-orbital interactions,
through the diagonal of $U_{mm'}^{\sigma\bar{\sigma}}|_{\textrm{cRPA}}$,
since the cubic symmetry is an approximation for the crystal field
in FeSe. The deviation is around 0.14 eV and the average intra-orbital
interaction (before shell-folding) calculated with cubic symmetry
is $U_{m}=7.06$ eV.

This deviation does not increase when elongating the crystal structure
through the c-direction, perpendicular to the Fe-Se tetrahedra, although
the dimensionality is reduced. This shows that the chemical environment
of Fe is the main actor for the accuracy of the Slater parametrization.

For the \emph{Slater symmetrized} reduced interaction matrices (in
eV), we get from equations \ref{Uslater1} and \ref{Uslater2}: 
\[
\bar{U}_{mm'}^{\sigma\sigma}|_{\textrm{Slater}}=\left(\begin{array}{ccccc}
0 & 2.55 & 2.55 & 3.40 & 3.40\\
2.55 & 0 & 3.69 & 2.84 & 2.84\\
2.55 & 3.69 & 0 & 2.84 & 2.84\\
3.40 & 2.84 & 2.84 & 0 & 2.84\\
3.40 & 2.84 & 2.84 & 2.84 & 0
\end{array}\right)
\]
 
\[
\bar{U}_{mm'}^{\sigma\bar{\sigma}}|_{\textrm{Slater}}=\left(\begin{array}{ccccc}
4.95 & 3.35 & 3.35 & 3.92 & 3.92\\
3.35 & 4.95 & 4.11 & 3.54 & 3.54\\
3.35 & 4.11 & 4.95 & 3.54 & 3.54\\
3.92 & 3.54 & 3.54 & 4.95 & 3.54\\
3.92 & 3.54 & 3.54 & 3.54 & 4.95
\end{array}\right).
\]
 The deviation from the directly calculated values is of the order
of $0.10$ eV with a relative error of around $2\%$. Within the Slater
parametrization the intra-orbital interactions are orbital-independent
(that is, $U_{mm}$ is independent of $m$).

Considering the higher-order Slater integrals, $F^{2}=7.57$ eV and
$F^{4}=5.30$ eV, the screened ratio $F^{4}/F^{2}=0.70$ deviates
from the empirical value of $0.63$ (see Table \ref{pnictide-1}).
Such deviation does not imply that the localized orbitals for Fe in
FeSe display orbital anisotropies since the atomic parametrization
appears to be well justified, as shown above. For the bare (unscreened)
ratio $F^{4}/F^{2}|_{\textrm{bare}}$ we recover the atomic value.

\subsubsection{LaFeAsO}

For LaFeAsO, we find -- again writing the matrix within the set of
orbitals $d_{3z^{2}-r^{2}}$,$d_{x^{2}-y^{2}}$,$d_{xy}$,$d_{xz}$,$d_{yz}$
-- and after shell-folding with the d-to-ligand interaction $U_{dp}=1.17$
eV:
\[
U_{mm'}^{\sigma\sigma}|_{\textrm{cRPA}}=\left(\begin{array}{ccccc}
0 & 0.94 & 0.85 & 1.53 & 1.53\\
0.94 & 0 & 1.82 & 1.12 & 1.12\\
0.85 & 1.82 & 0 & 1.00 & 1.00\\
1.53 & 1.12 & 1.00 & 0 & 1.00\\
1.53 & 1.12 & 1.00 & 1.00 & 0
\end{array}\right)
\]
 
\[
U_{mm'}^{\sigma\bar{\sigma}}|_{\textrm{cRPA}}=\left(\begin{array}{ccccc}
3.04 & 1.66 & 1.51 & 1.97 & 1.97\\
1.66 & 3.23 & 2.19 & 1.73 & 1.73\\
1.51 & 2.19 & 2.65 & 1.57 & 1.57\\
1.97 & 1.73 & 1.57 & 2.68 & 1.55\\
1.97 & 1.73 & 1.57 & 1.55 & 2.68
\end{array}\right),
\]

The intra-orbital repulsions are larger on the $d_{3z^{2}-r^{2}}$
and $d_{x^{2}-y^{2}}$ orbitals. This effect is due to the smaller
orbital spreads of these orbitals that do not point toward the As
ligands \cite{LaOFeAs-veronica}. The differences between the intra-orbital
interactions for different orbitals are larger than in FeSe. In particular,
the deviation yields 0.58 eV between the intra-orbital interactions
on $d_{x^{2}-y^{2}}$ and $d_{xy}$ orbitals, against 0.14 eV for
FeSe.

Within the Slater parametrization, the symmetrized reduced interaction
matrices read: 
\[
\bar{U}_{mm'}^{\sigma\sigma}|_{\textrm{Slater}}=\left(\begin{array}{ccccc}
0 & 0.87 & 0.87 & 1.55 & 1.55\\
0.87 & 0 & 1.77 & 1.10 & 1.10\\
0.87 & 1.77 & 0 & 1.10 & 1.10\\
1.55 & 1.10 & 1.10 & 0 & 1.10\\
1.55 & 1.10 & 1.10 & 1.10 & 0
\end{array}\right)
\]
 
\[
\bar{U}_{mm'}^{\sigma\bar{\sigma}}|_{\textrm{Slater}}=\left(\begin{array}{ccccc}
2.85 & 1.53 & 1.53 & 1.98 & 1.98\\
1.53 & 2.85 & 2.13 & 1.68 & 1.68\\
1.53 & 2.13 & 2.85 & 1.68 & 1.68\\
1.98 & 1.68 & 1.68 & 2.85 & 1.68\\
1.98 & 1.68 & 1.68 & 1.68 & 2.85
\end{array}\right).
\]
 The largest discrepancy with the direct calculation, around $0.38$
eV, is obtained for $d_{x^{2-}y^{2}}$. It is the hybridization with
the As ligands and the covalent character of the As-Fe bonding that
induce larger deviations from the atomic sphericity than in FeSe.
Se atoms have a Pauling electronegativity of around 2.55 that is larger
than the one of As (2.18) or Fe (1.83). The more ionic character of
the Fe-Se bonding makes the localized Fe-3\textit{d} orbitals more
atomic-like, and hence the Slater parametrization more accurate.

\subsubsection{BaFe$_{2}$As$_{2}$ and BaRu$_{2}$As$_{2}$}

Similar arguments can be employed for understanding the Slater parametrization
for BaFe$_{2}$As$_{2}$. The interaction matrices (shell-folded with
$U_{dp}=1.33$ eV) now read
\[
U_{mm'}^{\sigma\sigma}|_{\textrm{cRPA}}=\left(\begin{array}{ccccc}
0 & 1.22 & 1.15 & 1.84 & 1.86\\
1.22 & 0 & 2.15 & 1.36 & 1.37\\
1.15 & 2.15 & 0 & 1.29 & 1.30\\
1.84 & 1.36 & 1.29 & 0 & 1.27\\
1.86 & 1.37 & 1.30 & 1.27 & 0
\end{array}\right)
\]
 
\[
U_{mm'}^{\sigma\bar{\sigma}}|_{\textrm{cRPA}}=\left(\begin{array}{ccccc}
3.47 & 1.96 & 1.86 & 2.31 & 2.33\\
1.96 & 3.47 & 2.53 & 1.99 & 2.00\\
1.86 & 2.53 & 3.11 & 1.89 & 1.90\\
2.31 & 1.99 & 1.89 & 3.04 & 1.85\\
2.33 & 2.00 & 1.90 & 1.85 & 3.07
\end{array}\right).
\]
 whereas the \emph{Slater symmetrized} interaction matrices equal:
\[
\bar{U}_{mm'}^{\sigma\sigma}|_{\textrm{Slater}}=\left(\begin{array}{ccccc}
0 & 1.12 & 1.12 & 1.85 & 1.85\\
1.12 & 0 & 2.09 & 1.36 & 1.36\\
1.12 & 2.09 & 0 & 1.36 & 1.36\\
1.85 & 1.36 & 1.36 & 0 & 1.36\\
1.85 & 1.36 & 1.36 & 1.36 & 0
\end{array}\right)
\]
 
\[
\bar{U}_{mm'}^{\sigma\bar{\sigma}}|_{\textrm{Slater}}=\left(\begin{array}{ccccc}
3.22 & 1.82 & 1.82 & 2.30 & 2.30\\
1.82 & 3.22 & 2.47 & 1.98 & 1.98\\
1.82 & 2.47 & 3.22 & 1.98 & 1.98\\
2.30 & 1.98 & 1.98 & 3.22 & 1.98\\
2.30 & 1.98 & 1.98 & 1.98 & 3.22
\end{array}\right).
\]
 The largest discrepancy with the direct calculation is about $0.25$
eV and is obtained for the $d_{3z^{2}-r^{2}}$-orbital, which points
toward the interlayer Ba planes and for $d_{x^{2}-y^{2}}$.

The larger hybridization of Ru 4\textit{d} states with the As ligands
also makes the atomic-like Slater parametrization less accurate. We
note that the largest value of the screened ratio $\ensuremath{F^{4}/F^{2}}$
is obtained for this compound.

\subsubsection{LiFeAs}

The same procedure as above is applied to LiFeAs, where we find $U_{dp}=1.85$
eV. Of all studied pnictides compounds, LiFeAs is the closest to FeSe
with larger Coulomb interactions and lower screened ratio $\ensuremath{F^{4}/F^{2}}$.
We obtain for the interactions matrices after shell-folding:

\[
U_{mm'}^{\sigma\sigma}|_{\textrm{cRPA}}=\left(\begin{array}{ccccc}
0 & 1.84 & 1.82 & 2.64 & 2.64\\
1.84 & 0 & 2.84 & 2.05 & 2.05\\
1.82 & 2.84 & 0 & 2.02 & 2.03\\
2.64 & 2.05 & 2.02 & 0 & 2.05\\
2.64 & 2.05 & 2.03 & 2.05 & 0
\end{array}\right)
\]
 
\[
U_{mm'}^{\sigma\bar{\sigma}}|_{\textrm{cRPA}}=\left(\begin{array}{ccccc}
4.19 & 2.60 & 2.57 & 3.12 & 3.12\\
2.60 & 4.07 & 3.23 & 2.71 & 2.71\\
2.57 & 3.23 & 3.95 & 2.68 & 2.69\\
3.12 & 2.71 & 2.68 & 3.99 & 2.68\\
3.12 & 2.71 & 2.69 & 2.68 & 3.99
\end{array}\right).
\]

whereas for the \emph{Slater symmetrized} interaction matrices we
obtain: 
\[
\bar{U}_{mm'}^{\sigma\sigma}|_{\textrm{Slater}}=\left(\begin{array}{ccccc}
0 & 1.81 & 1.81 & 2.60 & 2.60\\
1.81 & 0 & 2.86 & 2.07 & 2.07\\
1.81 & 2.86 & 0 & 2.07 & 2.07\\
2.60 & 2.07 & 2.07 & 0 & 2.07\\
2.60 & 2.07 & 2.07 & 2.07 & 0
\end{array}\right)
\]
 
\[
\bar{U}_{mm'}^{\sigma\bar{\sigma}}|_{\textrm{Slater}}=\left(\begin{array}{ccccc}
4.04 & 2.55 & 2.55 & 3.08 & 3.08\\
2.55 & 4.04 & 3.25 & 2.73 & 2.73\\
2.55 & 3.25 & 4.04 & 2.73 & 2.73\\
3.08 & 2.73 & 2.73 & 4.04 & 2.73\\
3.08 & 2.73 & 2.73 & 2.73 & 4.04
\end{array}\right).
\]

The parametrization is better than for other pnictides: the maximum
discrepancy between the parametrized and the directly calculated matrices
is only 0.15 eV. This, again, is the sign of more atomic-like Wannier
functions, due to the larger Fe-As distance.

\section{Screening channels}

\label{sec:Screening channels}

In addition to the physically motivated Hubbard interactions -- to
be used in low-energy models for the respective compounds -- other
partially screened interactions can be constructed, with the aim of
analyzing the importance of different screening processes. To this
effect, we choose different particle-hole transitions (``screening
channels'') that are removed from the RPA polarization. The effects
of these screening channels are not additive, since the interaction
depends on these different partial polarizations in a highly non-linear
manner.

In this section, we calculate different partially screened interactions
at zero frequency in the iron pnictides and chalcogenides and in SrVO\textsubscript{3}.
A detailed comparison allows us to understand global trends and to
compare the relative importance of the screening contributions from
the ligand \textit{p} and from the \textit{d} orbitals.

\subsection{Global trends along the pnictides and chalcogenide series}

\begin{figure}
\centering{}\includegraphics[width=8cm]{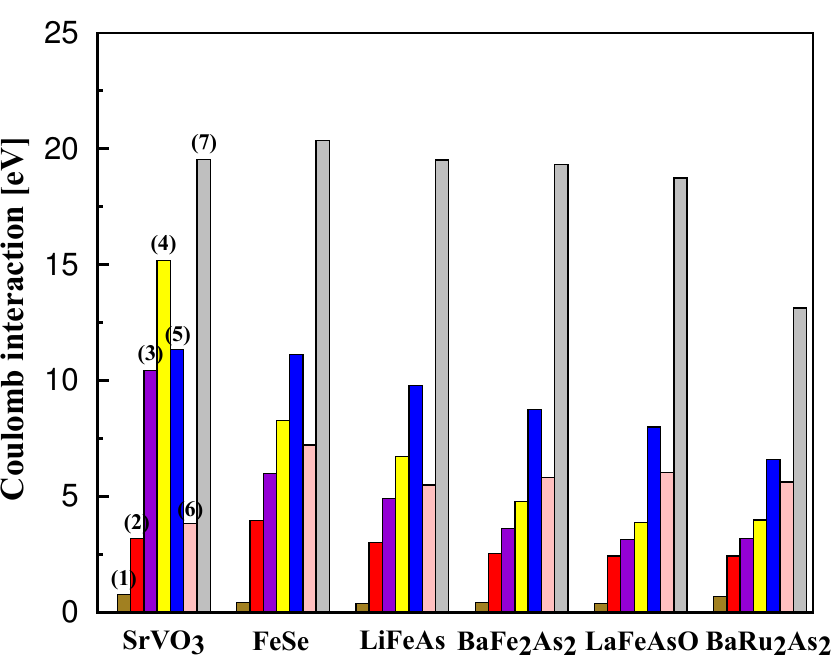} \caption[Strength of the screening channels in SrVO\textsubscript{3}, FeSe,
LiFeAs, BaFe$_{2}$As$_{2}$, LaFeAsO and BaRu$_{2}$As$_{2}$]{Strength of the screening channels in the comparison of an early
transition metal oxide (SrVO\textsubscript{{\footnotesize 3}}) with
iron-based pnictides. The bar charts show the static values of the
monopole part of partially screened interactions for the 3\textit{d}
local orbitals within the \textit{dp} low-energy Hamiltonians when
removing specific ``occupied to empty'' transitions from the total
RPA polarization. Cases (1) (brown) and (7) (gray) respectively correspond
to the fully screened and unscreened cases. Case (2) (red) is the
value of the static average intra-orbital interaction when removing
$d\rightarrow d$ transitions only (corresponding to what is commonly
denoted as \textit{d-dp} Hamiltonian in the literature \cite{cRPA-LaOFeAs-miyake}).
Case (3) (purple) corresponds to the average intra-orbital interaction
within the \textit{dp} Hamiltonian, i.e. removing all the transitions
within the energy window $\mathbb{W}_{dp}$, whereas in Case (4) (yellow),
all the transitions involving \textit{p} as well as $d\rightarrow d$
are removed. In Case (5) (blue), all the transitions involving \textit{d}
states are removed, whereas in Case (6) (salmon), only the transitions
involving empty \textit{d} states are considered.\label{histo} }
\end{figure}

For each compound, we calculate seven different quantities at zero
frequency, which are shown on Figure \ref{histo}. The bare values
on the local \textit{d} orbitals vary less than 10\% (except for the
case of BaRu\textsubscript{2}As\textsubscript{2} because of the
larger extension of the 4\textit{d} orbitals), as well as the fully
screened values. Thus one can directly compare the values of the intra-orbital
interactions obtained when considering specific occupied to empty
transitions. While SrVO\textsubscript{3} stands out, the structure
of the screening in all calculated pnictides and chalcogenide is remarkably
similar.

If we look at the relative importance of the channels, one main difference
between the different compounds is visible: the relative magnitude
of the interaction in case (4) where all the transitions involving
occupied \textit{p} and all \textit{d$\rightarrow$d} transitions
have been cut, and in case (6) where only the transitions from all
occupied states (except \textit{d}) to empty \textit{d} states are
considered. This is similar to comparing the screening of all occupied
except \textit{p} and \textit{d} to all empty states except \textit{d}
(case (4)) with the screening of \textit{p$\rightarrow$d} (case (6)).
In FeSe where there is no interlayer atom and LiFeAs where Li electrons
are deep core states, the number of channels of case (4) is reduced
compared to the cases of BaFe\textsubscript{2}As\textsubscript{2},
LaFeAsO and BaRu\textsubscript{2}As\textsubscript{2} where the interlayer
atoms provide more screening channels. That is why the interaction
in case (4) becomes bigger than in case (6) in FeSe and LiFeAs, while
it is the opposite for other compounds.

For the same reason, the values of partially screened interactions
are globally enhanced in FeSe and LiFeAs, because there are less possibilities
of transitions. That is also why LaFeAsO displays lower values of
the interaction. Eventually, the differences in screening within the
pnictides and chalcogenides family happen to be mostly due to the
interlayer structure.

\subsection{Screening contributions from the ligands p orbitals}

\subsubsection{Ligand p to d transitions}

To analyze these transitions we compare the values of cases (2) and
(3) of Figure \ref{histo}. The only difference between those two
cases is precisely that ligand \textit{p} to \textit{d} transitions
have been cut. In SrVO\textsubscript{3} the reduction of the Coulomb
interaction due to this channel is remarkable, about 63\%. Moreover,
by comparing to the bare value and to the fully screened value, we
see that these \textit{d$\rightarrow$p} transitions account for about
40\% of the total screening. On the other hand, in pnictides and chalcogenides
this reduction lies between 20\% (LaFeAsO and BaRu\textsubscript{2}As\textsubscript{2})
and 33\% (LiFeAs), two to three times less, and the \textit{d$\rightarrow$p}
transitions account for only 10\% of the total screening. This could
be expected given the large number of ligands surrounding the metal
in SrVO\textsubscript{3} and their ionic character. In the pnictides,
the As-Fe electronegativity difference is smaller, thus the bonding
is more covalent and the electrons are less free to rearrange their
density to screen the charge. Moreover, if we think in terms of transitions
from ligand filled to metal empty bands, there are simply more possibilities
in SrVO$_{3}$ than in iron pnictides. Indeed, due to the low filling
of the V-3\textit{d} shell with only one electron, nearly all ligand
to 3\textit{d} transitions contribute to the screening in SrVO3, while
in the pnictides the $d^{6}$ filling prevents most such transitions.

\subsubsection{Ligand p to other empty states}

Now we compare cases (3) and (4). In case (4) all transitions from
p states to other states than \textit{d} have been further suppressed.
Again we can see that while in SrVO\textsubscript{3} these transitions
give a reduction of 29\% of the Coulomb interaction, in the pnictides/chalcogenides
it is only a reduction of 17\% to 23 \%. The transitions from ligand
\textit{p} to other states are not as important as the transitions
to \textit{d} states, which could be expected since the \textit{d}
states are closer to the Fermi level.

\subsection{Screening contributions from d orbitals}

Let us examine cases (5) and (6). In case (5) all transitions involving
\textit{d} states have been removed, while in case (6) only transitions
to empty \textit{d} states are considered. The difference can tell
us how important the contribution of the \textit{d} orbitals to the
screening in the materials is.

\subsubsection{SrVO\textsubscript{3}}

In SrVO\textsubscript{3} we see that the transitions to empty \textit{d}
states are nearly enough to recover the value obtained within a \textit{d-dp}
calculation. In this oxide the main channels are related to the empty
\textit{d} states, and the \textit{p$\rightarrow$d} channel is predominant.
Still, suppressing all these channels allows to screen about 60\%
of the bare value.

\subsubsection{Pnictides and chalcogenides}

In the iron pnictides and chalcogenides the difference is not as impressive
as in SrVO\textsubscript{3}. In the extreme case of BaRu\textsubscript{2}As\textsubscript{2},
cases (5) and (6) nearly give the same result, showing that the empty
\textit{d} states are not as important. For 3\textit{d} compounds,
these transitions recover some predominance, and for FeSe and LiFeAs
we can see that they account for a large part of the screening of
the \textit{d-dp} model. This importance is reduced in materials with
interlayer screening atoms, as is also shown by the reduction of the
value of case (5). Finally, we see that in pnictides and chalcogenides
the main channels involve the \textit{d} states. Also, this family
is characterized by a very similar FeAs layer and the ligand \textit{p}
states are not dominant in the screening. This is why we attribute
the small differences in the screening of the Coulomb interactions
within the iron pnictides family to the interlayer structure. Indeed,
the atoms between layers can participate to the screening by adding
possibilities of transitions involving the \textit{d} states, and
the efficiency of these transitions depends on the material. We can
also see this effect from an atomic point of view and presume that
the higher polarizability of a large Ba ion in BaFe\textsubscript{2}As\textsubscript{2}
will be more efficient in screening the monopole interaction than
the smaller Li ion in LiFeAs.

\section{Frequency dependence}

\label{sec:Frequency dependence}

In this section, we discuss the frequency-dependence of the Hubbard
interactions in the iron pnictides and chalcogenides and in SrVO\textsubscript{3}.
We display the frequency-dependent intrashell and intershell interactions
in the \textit{dp-dp} Hamiltonian in which all transitions involving
the Fe-\textit{d} and As-\textit{p} orbitals are taken out. The structure
of the high-frequency tail is compared to the free electron-like plasmon
frequencies obtained for different numbers of electrons. We then calculate
the strength of the bosonic renormalization factor and the impact
of the shell-folding procedure on this quantity. Finally, we discuss
the screening of the multipole Slater integrals.

\subsection{High-frequency tail of the monopole interaction}

\begin{figure*}
\centering{}\includegraphics[bb=5bp 0bp 475bp 553bp,width=14cm]{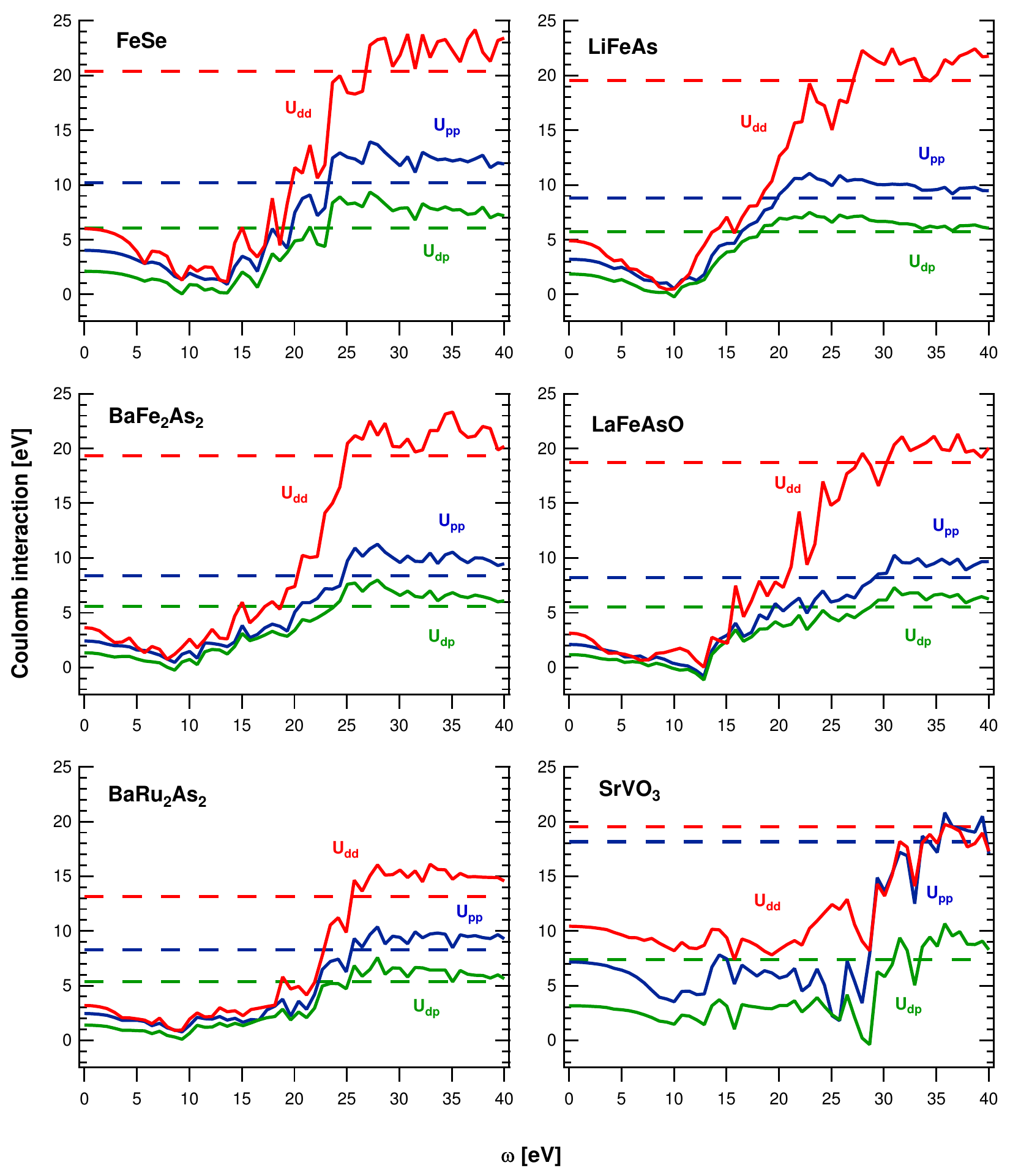}
\caption[Frequency-dependent Hubbard interactions of the \textit{dp} Hamiltonian
in SrVO\textsubscript{3}, FeSe, LiFeAs, BaFe$_{2}$As$_{2}$, LaFeAsO
and BaRu$_{2}$As$_{2}$]{Frequency dependence of the different Hubbard interactions in the
\textit{dp} Hamiltonian without shell-folding. Here $U_{dd}=F^{0}$
and $U_{pp}$ should be understood as the monopole part of the partially
screened interactions. 
Dashed lines are the values at infinite frequency.\label{dp_Hamiltonian}}
\end{figure*}

We calculate the real part of the Slater integrals as a function of
real frequency in the pnictides and chalcogenides. The same procedure
is also applied to SrVO$_{3}$ to use this compound as a benchmark.
A complete view of the \textit{dp-dp} Hamiltonian before shell-folding
is shown in Figure \ref{dp_Hamiltonian}.

For non-entangled systems where the hybridization between the correlated
atom and the ligand is small, the \textit{d} bands can be clearly
defined and separated from the ligand bands. As a consequence, calculating
the strength of the static Coulomb interaction by shell folding of
a \textit{dp-dp} Hamiltonian or using a \textit{d-dp} scheme where
only transitions from and to bands with a majority of \textit{d}-orbital
character are cut and $U_{dp}$ is neglected will give about the same
result. That is the case in most iron pnictides because the entanglement
is still relatively small and the \textit{d} bands can be reasonably
defined.

However, this is not true anymore if we look at the bare value. Indeed,
the bare repulsion is essentially related to the spread of the Wannier
function of the correlated orbital since no screening processes happen.
So before shell-folding the strength of the Coulomb interaction within
the \textit{d-}shell is the same for both \textit{dp-dp} and \textit{d-dp}
calculations. However there is a big difference in the treatment of
the intershell interaction $U_{dp}$. While in a \textit{dp-dp} model
we see that the bare value of $U_{dp}$ is much larger than the zero-frequency
value, it is just ignored in a \textit{d-dp} ``entangled'' calculation.

This motivates the application of frequency-dependent shell-folding.
The results for all studied pnictides and chalcogenide are displayed
on Figure \ref{Real_F0_dp_vs_d-dp}, and compared to an ``entangled''
\textit{d-dp} calculation where only transitions from and to bands
with a majority of \textit{d}-orbital character are cut.

For all compounds, at first sight the main correction introduced by
the effective model is on the high-frequency tail, while the static
part stays essentially the same. The infinite frequency value is lowered
by about 30\%. On the other hand, the frequency dependence of both
models looks similar, with peaks around the same values of $\omega$.
Those peaks are less sharp in the effective model. Indeed $F_{0}$
and $U_{dp}$ also share the same frequency dependence so $U_{eff}=F_{0}-U_{dp}$
is smoothened compared to $F_{0}$.

Interestingly, the case of SrVO\textsubscript{3} is much different
from the pnictides, since both $U(0)$ and the high frequency tail
are substantially modified when we take $U_{dp}$ into consideration.
Moreover, while in pnictides taking the \textit{p-d} interactions
into account leaves $U(0)$ stable or reduces it, in SrVO\textsubscript{3}
$U(0)$ is increased. The correction of this value induced by the
effective model seems to be in agreement with values used in many-body
calculations where all \textit{d}-orbitals are taken into account.

\subsection{Plasmons and interband transitions}

\begin{figure*}
\centering{}\includegraphics[bb=5bp 0bp 485bp 553bp,width=14cm]{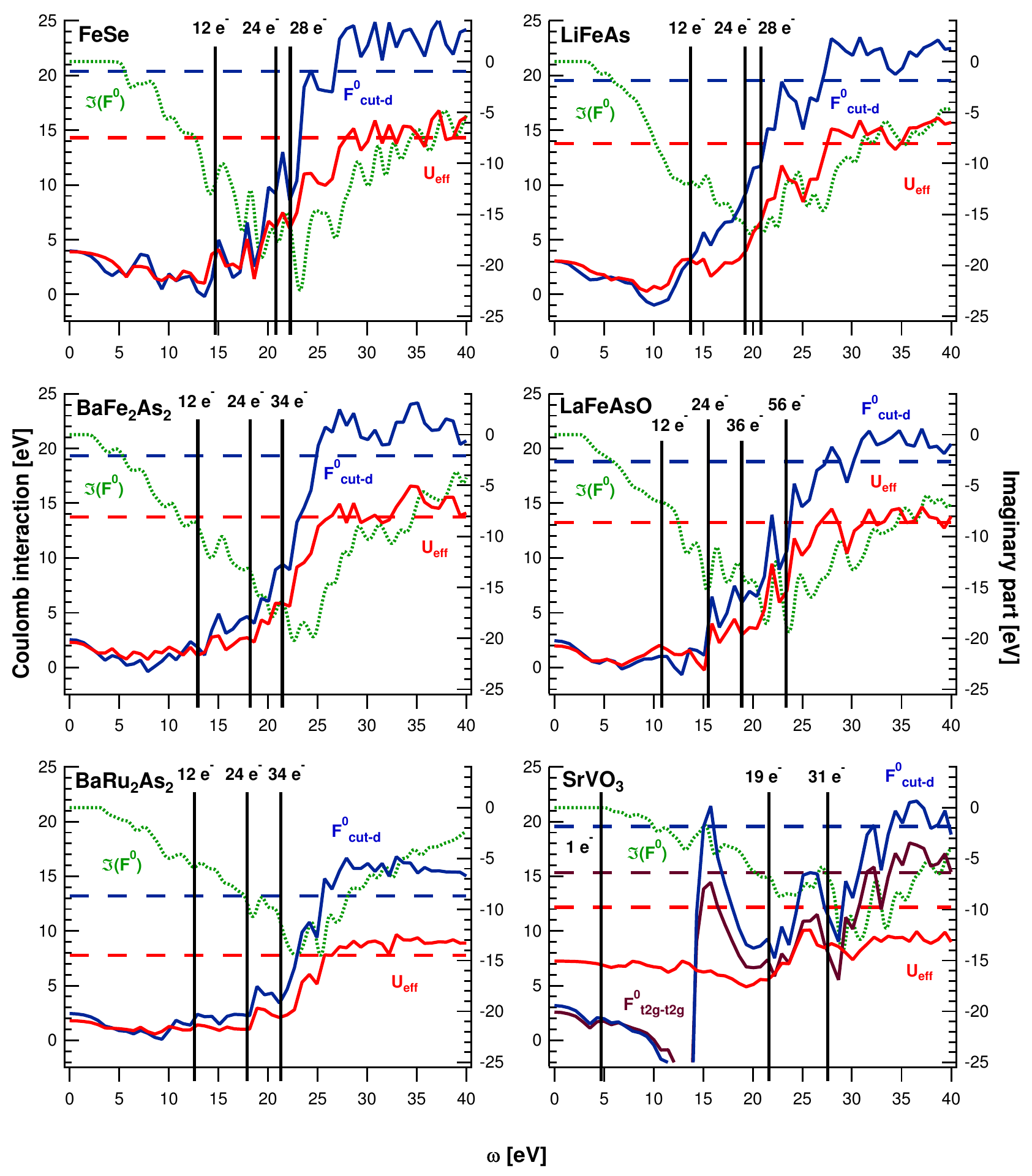}
\caption[Frequency dependence of the monopole part of the Hubbard interaction
in SrVO\textsubscript{3}, FeSe, LiFeAs, BaFe$_{2}$As$_{2}$, LaFeAsO
and BaRu$_{2}$As$_{2}$ in a \textit{d-dp} model with and without
shell-folding]{Frequency dependence of the monopole part of the partially screened
Coulomb interaction within the \textit{d}-shell. $U_{eff}(\omega)$
refers to $F^{0}(\omega)-U_{dp}(\omega)$ calculated in the \textit{dp-dp}
model. $F_{cut-d}^{0}$ is the monopole part of the interaction calculated
in an ``entangled'' \textit{d-dp} model where only $d\rightarrow d$
transitions are removed. Dashed lines are the values at infinite frequency.
The dotted line corresponds to the imaginary part of $F^{0}(\omega)$
and is compared to a free-electron calculation of the plasma frequency
at partial resonances (vertical bars).\label{Real_F0_dp_vs_d-dp}}
\end{figure*}

The dynamical structure of the Coulomb interaction is directly linked
to the variations of the constrained polarization $P^{r}$. These
variations are determined by interband transitions and collective
excitations. Indeed, in iron pnictides, ion-core polarization can
be neglected, as shown by calculations of the constrained macroscopic
dielectric function where all transitions from and to the valence
electron bands have been cut. We calculate the energy of the main
plasmon mode based on the free-electron formula for the plasma frequency:
\begin{equation}
\omega_{p}=\sqrt{\frac{ne^{2}}{m\epsilon_{0}}}
\end{equation}

To calculate the density we take into account all valence electrons,
which corresponds to all bands down to -20 eV: the Fe 3\textit{d}
electrons, As or Se 4\textit{p} and 4\textit{s} electrons, Ba 4\textit{p}
in BaFe\textsubscript{2}As\textsubscript{2} and BaRu\textsubscript{2}As\textsubscript{2},
and La 5\textit{p}, O 2\textit{p} and 2\textit{s} in LaFeAsO. The
binding energies of those electrons are still lower than the obtained
plasma frequency (between 20 and 25 eV), so it is reasonable to think
that they will enter the collective resonance. For SrVO\textsubscript{3}
we take into account V 3\textit{d}, O 2\textit{p}, Sr 4\textit{p}
and O 2\textit{s} electrons. The plasmon frequency is then compared
to the imaginary part of the monopole interaction in a \textit{dp}
model (see the plasmon energy corresponding to the highest number
of electrons for each compound on Figure \ref{Real_F0_dp_vs_d-dp}).
We can see that the main peak of $\Im(F^{0})$ agrees very well with
the calculated plasmon frequency and corresponds to a cutoff frequency
where the Coulomb interaction increases sharply from the static value
to the infinite frequency value.

We also show that some of the other peaks could be assigned to partial
plasmon resonances. The first partial resonance would correspond to
the number of Fe 3\textit{d} (or V 3\textit{d}) electrons. For the
second one, we add the As or Se 4\textit{p} electrons (or the O 2\textit{p}
in the case of SrVO\textsubscript{3}). We also show a third partial
resonance in LaFeAsO, corresponding to the addition of the O 2\textit{p}
electrons. However, it is difficult to make a one-to-one correspondence
because of the entanglement of the plasmons with the interband transitions
which creates rather a continuum of screening modes. This effect has
been documented in transition metals, where the interband transitions
from the \textit{d} bands to higher bands act to shift and broaden
the plasmons composed of \textit{s} and \textit{p} electrons \cite{pines-plasmons}.
It also happens in the pnictides and chalcogenides since there are
possibilities of transitions at frequencies close to the one of the
free-electron plasmon. As an illustration, we can see in the case
of SrVO\textsubscript{3} that the agreement between a partial resonance
of O 2\textit{p} and V 3\textit{d} electrons with the plasmon around
15 eV is really bad. However, the free-electron plasma frequency of
about 21.5 eV is likely to be modified by a combination of possible
interband transitions in this frequency range and background polarization
provided by lower-lying states. Indeed, the calculation of the constrained
macroscopic dielectric function where transitions from \textit{dp}
bands to all empty states have been removed gives a value of about
1.4 at $\omega=16$ eV. Simply taking this background macroscopic
dielectric function into account would already reduce the plasma frequency
to around 18 eV. Further adding the contribution of the possible interband
transitions could easily shift the value of the plasma frequency to
15 eV.

\subsection{Density of screening modes}

\begin{figure*}
\centering{}\includegraphics[bb=5bp 0bp 475bp 446bp,width=14cm]{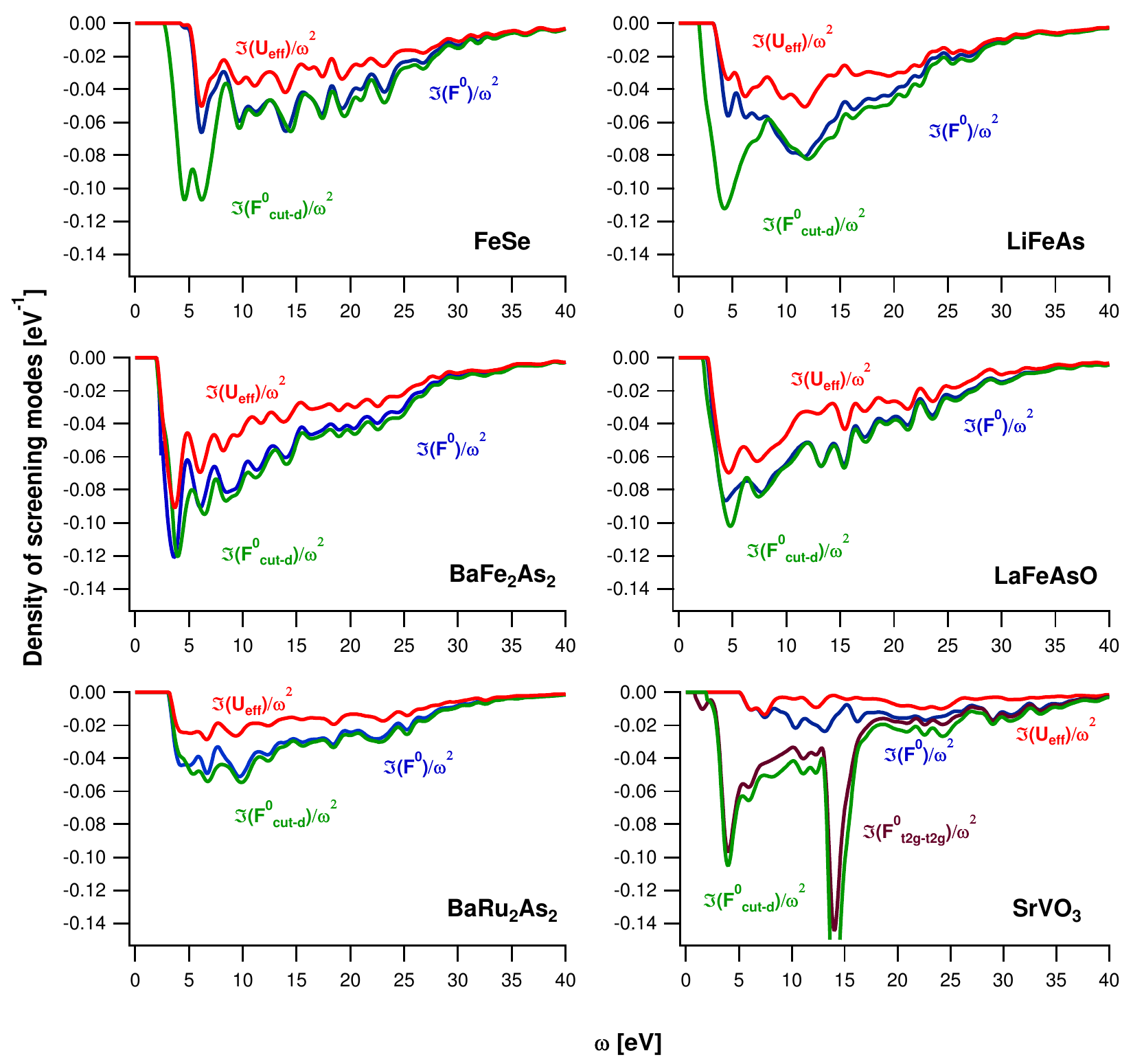}
\caption[Density of screening modes in SrVO\textsubscript{3}, FeSe, LiFeAs,
BaFe$_{2}$As$_{2}$, LaFeAsO and BaRu$_{2}$As$_{2}$ in a \textit{d-dp}
model with and without shell-folding]{Density of screening modes in iron pnictides and chalcogenides before
and after shell folding of a \textit{dp-dp }model compared to an entangled
\textit{d-dp} model. SrVO\textsubscript{{\footnotesize 3}} is also
shown for comparison.\label{Screening_modes}}
\end{figure*}

We can now examine the impact of taking into account the \textit{d-p}
interaction on $\Im(U(\omega))/\omega^{2}$, which can be physically
understood as the density of screening modes and determines $Z_{B}$
(see Figure \ref{Screening_modes}). Though the global structure is
conserved, a strong renormalization is induced.

Two effects are successively involved. First, we cut more transitions
in the cRPA calculation in the \textit{dp-dp} model. This will have
an impact at low frequency, especially if the first transitions happening
in the \textit{d-dp} model were from occupied \textit{p} to empty
\textit{d}. In that case the gap of the screening modes, that is to
say the energy needed for the first transition between occupied and
empty bands, will increase. At higher frequencies, the difference
is largely negligible. Indeed, most of the screening processes do
not involve \textit{p$\rightarrow$d} transitions, as shown in Section
\ref{sec:Screening channels}. The second effect is due to the shell-folding
procedure. Due to the fact that $U_{dd}$ and $U_{dp}$ share variations
in frequency, the frequency dependence of the effective interaction
is flattened. Eventually, most of the dynamical structure of the screening
stays unchanged when we suppress \textit{p$\rightarrow$d} transitions,
and the main effect of shell-folding is a reduction of the density
of screening modes.

One could wonder about the validity of the different models depending
on the frequency. At very low frequency (below the gap of the \textit{d-dp}
model), the \textit{d-dp} model result is adapted for the pnictides,
because the entanglement is not too strong and the bands can be relatively
well separated. However, as soon as the frequency becomes larger than
the gap, this low-energy model is not valid anymore in the sense of
the renormalization group. For higher frequencies, this is all the
more true since the Coulomb interaction between \textit{d} and \textit{p}
orbitals is even higher and cannot be neglected.

\subsection{Bosonic renormalization factor Z\textsubscript{B}}

We focus on the impact of the shell-folding procedure on the bosonic
renormalization factor $Z_{B}$ introduced in Section \ref{sub:Frequency dependence}.
The smaller density of screening modes in the effective model lowers
the value of $Z_{B}$ since 
\begin{equation}
\ln\left(Z_{B}\right)=-\int_{0}^{+\infty}\frac{\Im U_{ret}(\omega)}{\pi\omega^{2}}d\omega
\end{equation}
 We compare the values obtained for a \textit{d-dp} model with and
without shell-folding in Table \ref{ZB}.

At low frequency, the spectral weight reduction of the quasiparticles
is $Z_{eff}\times Z_{B}$ where $Z_{eff}$ is the renormalization
obtained in a static Hubbard model \cite{udyneff-michele}. $Z_{eff}$
depends on $U(0)$/D with D the bandwidth. In the pnictides case,
$U(0)$ is nearly the same in the two calculations. Qualitatively,
$Z_{eff}$ will be a little bigger in the effective model since both
$U(0)$ is slightly smaller and the bandwidth is larger due to a higher
$Z_{B}$. So the discrepancies between the two models as to the physical
properties of the system will be largely dependent on the bosonic
renormalization factor $Z_{B}$. Eventually, the renormalization is
substantially changed when we take into account the \textit{p-d} interactions.

As for SrVO\textsubscript{3}, the bosonic renormalization factor
$Z_{B}$ in the effective model is equal to 0.93, which should be
compared to a value of 0.70 in a \textit{t\textsubscript{\textit{2g}}-t\textsubscript{\textit{2g}}}
model (0.64 when cutting only the transitions from and to the \textit{d}-like
bands). We see that in this compound $U_{eff}$ is very close to being
static, while in the pnictides there is still a large frequency dependence
even in the effective model. Indeed, $U_{eff}$ is less screened in
SrVO\textsubscript{3}: the infinite frequency value is reduced by
40\%, while in the pnictides the reduction is much higher, from 70\%
in FeSe to 85\% in LaFeAsO. This indicates a low coupling to the plasmon.
The reason is that the main plasmon around 15 eV in SrVO\textsubscript{3}
is due to transitions from occupied O-\textit{p} to empty V-\textit{eg}
states (see Figures \ref{Screening_modes} and \ref{histo}), which
are suppressed in a \textit{dp-dp} model. The shell-folding procedure
allows us to take into account the intershell \textit{p-d} interaction
that was ignored in an entangled \textit{d-dp} model. However, when
we project our \textit{dp-dp} model into an effective \textit{d-dp}
model we lose the possibility to reintroduce the screening from p$\longrightarrow$d
transitions -- which might also necessitate a more refined model including
long-range interactions. We mention in particular that the frequency-dependence
of the effective local interaction is expected to become stronger
when non-local screening processes within the low-energy manifold
are taken into account. This would correspond to a generalisation
of what has been worked out in \cite{Nomura_local_U} for the static
part of the effective interactions. These arguments demonstrate that
the choice of the appropriate low-energy Hamiltonian remains a subtle
and crucial question, since the effects contained within the different
models are not the same.

\begin{table}
\caption[Values of $Z_{B}$ in FeSe, LiFeAs, BaFe$_{2}$As$_{2}$, LaFeAsO
and BaRu$_{2}$As$_{2}$ with and without shell-folding]{Values of $Z_{B}$ extracted from the monopole part of the interaction
within the \textit{d}-shell. Results for the shell-folded \textit{dp-dp}
model and for the directly calculated \textit{d-dp} model are displayed.\label{ZB}}

\centering{}{\footnotesize }%
\begin{tabular*}{1\columnwidth}{@{\extracolsep{\fill}}lccccc}
\toprule 
\toprule[0.08em]  & FeSe & LiFeAs & BaFe$_{2}$As$_{2}$ & LaFeAsO & BaRu$_{2}$As$_{2}$\tabularnewline
\midrule 
\midrule[0.05em] $Z_{B}$ (effective){\footnotesize{} } & 0.78 & 0.76 & 0.71 & 0.73 & 0.85\tabularnewline
$Z_{B}$ (entangled){\footnotesize{} } & 0.63 & 0.60 & 0.59 & 0.62 & 0.74\tabularnewline\bottomrule[0.08em] 
\bottomrule
\end{tabular*}
\end{table}

\subsection{Dynamical J}

We will now focus on the Hund's coupling matrix 
\begin{equation}
J_{mm'}=U_{mm'm'm,m\neq m'}
\end{equation}
 In a cubic basis we can define\textbf{:} 
\begin{equation}
\bar{J}\equiv\frac{5}{7}\frac{F^{2}+F^{4}}{14}
\end{equation}
 which physically corresponds to an arithmetic mean of all elements
$J_{mm'}$. This quantity is frequency-dependent and differs by a
factor 5/7 from the definition of Section \ref{sub:Slater parametrization}
-- which is more adapted to the case of a model defined only by $U$
and $J$, while here we are considering the full orbital-dependent
matrix. In BaFe\textsubscript{2}As\textsubscript{2}, it will vary
by about 17\%, from 0.58 eV at zero frequency to 0.68 eV at infinite
frequency. However, if one looks at the full $J$ matrix at zero frequency,
we find: 
\[
J_{mm'}|_{\textrm{cRPA}}=\left(\begin{array}{ccccc}
0 & 0.74 & 0.71 & 0.47 & 0.47\\
0.74 & 0 & 0.38 & 0.63 & 0.63\\
0.71 & 0.38 & 0 & 0.60 & 0.60\\
0.47 & 0.63 & 0.60 & 0 & 0.58\\
0.47 & 0.63 & 0.60 & 0.58 & 0
\end{array}\right).
\]


where the order of the orbitals is, as before, $d_{3z^{2}-r^{2}}$,$d_{x^{2}-y^{2}}$,$d_{xy}$,$d_{xz}$,$d_{yz}$.

We also give the corresponding bare $J$ 
\[
J_{mm'}|_{\textrm{cRPA}}=\left(\begin{array}{ccccc}
0.0 & 0.88 & 0.86 & 0.52 & 0.52\\
0.88 & 0.0 & 0.40 & 0.74 & 0.74\\
0.86 & 0.40 & 0.0 & 0.73 & 0.73\\
0.52 & 0.74 & 0.73 & 0.0 & 0.71\\
0.52 & 0.74 & 0.73 & 0.71 & 0.0
\end{array}\right)
\]

and the Slater-parametrized version of the low-frequency $J$: 
\[
J_{mm'}|_{\textrm{cRPA}}=\left(\begin{array}{ccccc}
0.0 & 0.70 & 0.70 & 0.45 & 0.45\\
0.70 & 0.0 & 0.38 & 0.62 & 0.62\\
0.70 & 0.38 & 0.0 & 0.62 & 0.62\\
0.45 & 0.62 & 0.62 & 0.0 & 0.62\\
0.45 & 0.62 & 0.62 & 0.62 & 0.0
\end{array}\right)
\]

\begin{figure*}
\centering{}\includegraphics[bb=5bp 0bp 490bp 553bp,width=14cm]{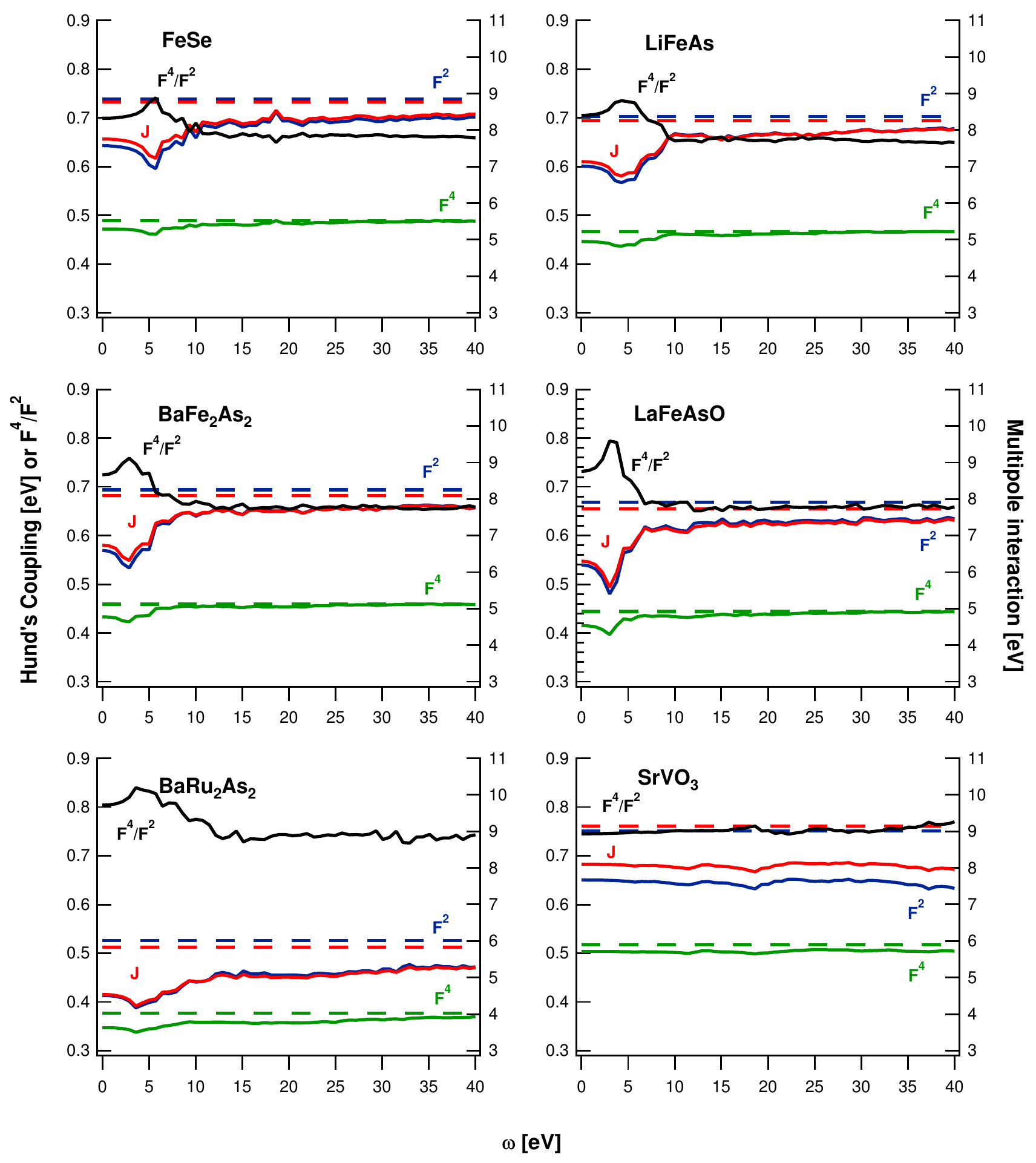}
\caption[Frequency dependence of $F^{2}$, $F^{4}$ and $J$ calculated in
a \textit{dp-dp} model in SrVO\textsubscript{3}, FeSe, LiFeAs, BaFe$_{2}$As$_{2}$,
LaFeAsO and BaRu$_{2}$As$_{2}$]{Frequency dependence of $F^{2}$, $F^{4}$ and $\bar{J}$$\left(\equiv\frac{5}{7}\frac{F^{2}+F^{4}}{14}\right)$
calculated in a \textit{dp-dp} model. Dashed lines are the values
at infinite frequency. Note the different energy scales for the Slater
integrals (right scale) and for $\bar{J}$ (left scale), in eV. Also
shown is the unitless ratio $F^{4}/F^{2}$, using the left scale,
but of course in this case as a dimensionless scale. \label{Multipole}}
\end{figure*}

The spread of the elements of these matrices is really large: for
instance, the $d_{z^{2}}\longleftrightarrow d_{x^{2}-y^{2}}$ element
is about twice the $d_{x^{2}-y^{2}}\longleftrightarrow d_{xy}$ element,
in all three matrices, demonstrating that this is a consequence of
the different orbital extensions. Consequently, $J$ is not a good
quantity to focus on, and it is better to look at the frequency-dependence
of the Slater integrals $F^{2}$ and $F^{4}$. This is shown on Figure
\ref{Multipole}. While $F^{4}$ shows very little variation with
$\omega$, $F^{2}$ exhibits a minimum at an intermediate frequency
which corresponds to the onset of interband transitions. It is also
a minimum of $J$ and a maximum of $F^{4}/F^{2}$.

In a DMFT calculation, one can wonder how to deal with this non-monopole
frequency dependence. We suggest several answers. Using the bare value
for $F^{2}$ and $F^{4}$ is satisfying from a model point of view.
At infinite frequency the Slater parametrization of the Coulomb interaction
matrix is excellent because the system is atomic like. Then one can
assume that the plasmon is only screening the monopole part of the
interaction. On the other hand, the low-energy properties of the system
are more influenced by the static part of the interaction. In this
view, the best would be to parametrize the Coulomb interaction matrix
at low frequency as well as possible, and then to ignore again the
effects of the plasmons on the non-monopole terms. Finally, one could
also consider a fully frequency-dependent matrix, using for example
the double expansion algorithm of Steiner et al. \cite{Steiner_double_expansion}.

\section{Conclusion}

In conclusion, we have studied the strength of the effective Hubbard
interactions and the accuracy of the Slater parametrization in the
iron pnictides for a shell-folded \textit{dp-dp} model. In agreement
to what was found in Ref. \onlinecite{loig-tesis} for the \textit{d-dp}
model, we find that the effective Coulomb interactions for Fe-3\textit{d}
shells are larger in 11 than in 122 and 1111 pnictides, while the
111 pnictides are an intermediate case, and that the accuracy of the
Slater parametrization depends on the ligand-metal bonding character
rather than on the dimensionality of the lattice: it is excellent
for ionic-like FeSe and not as good for more covalent Fe-As (LaFeAsO,
BaFe\textsubscript{2}As\textsubscript{2}) pnictides. The main effect
of the shell-folding procedure is to reduce the value of the high-frequency
value of the Coulomb interaction by about 30\%.

We have discussed the relative importance of screening channels which
reduce the on-site bare interaction to the fully screened one. We
have shown that the screening channels are analogously structured
in the pnictides and chalcogenides family, while this structure is
very different in the benchmark oxide SrVO\textsubscript{3}. The
ligand channel does not appear to be responsible for the dominant
screening mechanism in iron pnictides.

We have calculated the full frequency dependence of the Hubbard interaction
in the 11, 111, 122 and 1111 families of iron pnictides and compared
it to SrVO\textsubscript{3} in the \textit{dp-dp} model, including
both Fe-\textit{d} and As-\textit{p} degrees of freedom. We have calculated
the free-electron plasma frequencies corresponding to different numbers
of electrons involved in the resonance, and we have shown that the
screening modes could not be approximated by a single plasmon as in
SrVO\textsubscript{3}.

Finally, we have studied the effect of the shell-folding procedure
and compared the so-constructed effective \textit{d-dp} model to a
\textit{d-dp} model where only transitions from and to bands with
a majority of \textit{d}-orbital character are cut and $U_{dp}$ is
neglected. We find an important reduction of the high-frequency tail
which results in a less important bosonic renormalization factor $Z_{B}$
(that is, closer to one).

\section*{ACKNOWLEDGMENTS}

We acknowledge useful discussions with M. Aichhorn, F. Aryasetiawan,
M. Casula, A. Georges, C. Martins, T. Miyake and G. Sawatzky. This
work was supported by the European Research Council under its Consolidator
grant scheme (project number 617196), IDRIS/GENCI under project t2016091393,
the Natural Science Foundation of China (Projects No. 21211130098,
21373017 and 21321001) and the Cai Yuanpei program.

\bibliography{BiblioFirstprinciples,BiblioU,Biblio113,Biblio214,BiblioDFT,BiblioWannier,BiblioGeneral,BiblioDMFT,BiblioGW,Bibliopnictides,biblio_project,ListPublications}

\begin{thebibliography}{81}
\expandafter\ifx\csname natexlab\endcsname\relax\def\natexlab#1{#1}\fi
\expandafter\ifx\csname bibnamefont\endcsname\relax
  \def\bibnamefont#1{#1}\fi
\expandafter\ifx\csname bibfnamefont\endcsname\relax
  \def\bibfnamefont#1{#1}\fi
\expandafter\ifx\csname citenamefont\endcsname\relax
  \def\citenamefont#1{#1}\fi
\expandafter\ifx\csname url\endcsname\relax
  \def\url#1{\texttt{#1}}\fi
\expandafter\ifx\csname urlprefix\endcsname\relax\def\urlprefix{URL }\fi
\providecommand{\bibinfo}[2]{#2}
\providecommand{\eprint}[2][]{\url{#2}}

\bibitem[{\citenamefont{Hirschfeld}(2016)}]{Hirschfeld-CRAS}
\bibinfo{author}{\bibfnamefont{P.~J.} \bibnamefont{Hirschfeld}},
  \bibinfo{journal}{Comptes Rendus Physique} \textbf{\bibinfo{volume}{17}},
  \bibinfo{pages}{197 } (\bibinfo{year}{2016}), ISSN \bibinfo{issn}{1631-0705},
  \bibinfo{note}{iron-based superconductors / Supraconducteurs \`a base de
  fer}.

\bibitem[{\citenamefont{Rullier-Albenque}(2016)}]{Albenque-CRAS}
\bibinfo{author}{\bibfnamefont{F.}~\bibnamefont{Rullier-Albenque}},
  \bibinfo{journal}{Comptes Rendus Physique} \textbf{\bibinfo{volume}{17}},
  \bibinfo{pages}{164 } (\bibinfo{year}{2016}), ISSN \bibinfo{issn}{1631-0705},
  \bibinfo{note}{iron-based superconductors / Supraconducteurs \`a base de
  fer}.

\bibitem[{\citenamefont{van Roekeghem et~al.}(2016)\citenamefont{van Roekeghem,
  Richard, Ding, and Biermann}}]{Ambroise-CRAS}
\bibinfo{author}{\bibfnamefont{A.}~\bibnamefont{van Roekeghem}},
  \bibinfo{author}{\bibfnamefont{P.}~\bibnamefont{Richard}},
  \bibinfo{author}{\bibfnamefont{H.}~\bibnamefont{Ding}}, \bibnamefont{and}
  \bibinfo{author}{\bibfnamefont{S.}~\bibnamefont{Biermann}},
  \bibinfo{journal}{Comptes Rendus Physique} \textbf{\bibinfo{volume}{17}},
  \bibinfo{pages}{140 } (\bibinfo{year}{2016}), ISSN \bibinfo{issn}{1631-0705},
  \bibinfo{note}{iron-based superconductors / Supraconducteurs \`a base de
  fer}.

\bibitem[{\citenamefont{Kamihara et~al.}(2008)\citenamefont{Kamihara, Watanabe,
  Hirano, and Hosono}}]{Kamihara-2008}
\bibinfo{author}{\bibfnamefont{Y.}~\bibnamefont{Kamihara}},
  \bibinfo{author}{\bibfnamefont{T.}~\bibnamefont{Watanabe}},
  \bibinfo{author}{\bibfnamefont{M.}~\bibnamefont{Hirano}}, \bibnamefont{and}
  \bibinfo{author}{\bibfnamefont{H.}~\bibnamefont{Hosono}},
  \bibinfo{journal}{J. Am. Chem. Soc.} \textbf{\bibinfo{volume}{130}},
  \bibinfo{pages}{3296} (\bibinfo{year}{2008}).

\bibitem[{\citenamefont{Ren et~al.}(2008)\citenamefont{Ren, Wei, Jie, Wei,
  Xiao-Li, Zheng-Cai, Guang-Can, Xiao-Li, Li-Ling, Fang et~al.}}]{Ren-2008}
\bibinfo{author}{\bibfnamefont{Z.-A.} \bibnamefont{Ren}},
  \bibinfo{author}{\bibfnamefont{L.}~\bibnamefont{Wei}},
  \bibinfo{author}{\bibfnamefont{Y.}~\bibnamefont{Jie}},
  \bibinfo{author}{\bibfnamefont{Y.}~\bibnamefont{Wei}},
  \bibinfo{author}{\bibfnamefont{S.}~\bibnamefont{Xiao-Li}},
  \bibinfo{author}{\bibnamefont{Zheng-Cai}},
  \bibinfo{author}{\bibfnamefont{C.}~\bibnamefont{Guang-Can}},
  \bibinfo{author}{\bibfnamefont{D.}~\bibnamefont{Xiao-Li}},
  \bibinfo{author}{\bibfnamefont{S.}~\bibnamefont{Li-Ling}},
  \bibinfo{author}{\bibfnamefont{Z.}~\bibnamefont{Fang}}, \bibnamefont{et~al.},
  \bibinfo{journal}{Chin. Phys. Lett.} \textbf{\bibinfo{volume}{25}},
  \bibinfo{pages}{2215} (\bibinfo{year}{2008}).

\bibitem[{\citenamefont{Chen et~al.}(2008{\natexlab{a}})\citenamefont{Chen, Li,
  Wu, Li, Hu, Dong, Zheng, Luo, and Wang}}]{CeOFeAs-chen}
\bibinfo{author}{\bibfnamefont{G.~F.} \bibnamefont{Chen}},
  \bibinfo{author}{\bibfnamefont{Z.}~\bibnamefont{Li}},
  \bibinfo{author}{\bibfnamefont{D.}~\bibnamefont{Wu}},
  \bibinfo{author}{\bibfnamefont{G.}~\bibnamefont{Li}},
  \bibinfo{author}{\bibfnamefont{W.~Z.} \bibnamefont{Hu}},
  \bibinfo{author}{\bibfnamefont{J.}~\bibnamefont{Dong}},
  \bibinfo{author}{\bibfnamefont{P.}~\bibnamefont{Zheng}},
  \bibinfo{author}{\bibfnamefont{J.~L.} \bibnamefont{Luo}}, \bibnamefont{and}
  \bibinfo{author}{\bibfnamefont{N.~L.} \bibnamefont{Wang}},
  \bibinfo{journal}{Phys. Rev. Lett.} \textbf{\bibinfo{volume}{100}},
  \bibinfo{pages}{247002} (\bibinfo{year}{2008}{\natexlab{a}}).

\bibitem[{\citenamefont{Wen et~al.}(2008)\citenamefont{Wen, Mu, Fang, Yang, and
  Zhu}}]{LaOFeAs-Wen}
\bibinfo{author}{\bibfnamefont{H.-H.} \bibnamefont{Wen}},
  \bibinfo{author}{\bibfnamefont{G.}~\bibnamefont{Mu}},
  \bibinfo{author}{\bibfnamefont{L.}~\bibnamefont{Fang}},
  \bibinfo{author}{\bibfnamefont{H.}~\bibnamefont{Yang}}, \bibnamefont{and}
  \bibinfo{author}{\bibfnamefont{X.}~\bibnamefont{Zhu}}, \bibinfo{journal}{Eur.
  Phys. Lett.} \textbf{\bibinfo{volume}{82}}, \bibinfo{pages}{17009}
  (\bibinfo{year}{2008}).

\bibitem[{\citenamefont{Chen et~al.}(2008{\natexlab{b}})\citenamefont{Chen, Wu,
  Wu, Liu, Chen, and Fang}}]{SmOFeAs-Chen}
\bibinfo{author}{\bibfnamefont{X.~H.} \bibnamefont{Chen}},
  \bibinfo{author}{\bibfnamefont{T.}~\bibnamefont{Wu}},
  \bibinfo{author}{\bibfnamefont{G.}~\bibnamefont{Wu}},
  \bibinfo{author}{\bibfnamefont{R.~H.} \bibnamefont{Liu}},
  \bibinfo{author}{\bibfnamefont{H.}~\bibnamefont{Chen}}, \bibnamefont{and}
  \bibinfo{author}{\bibfnamefont{D.~F.} \bibnamefont{Fang}},
  \bibinfo{journal}{Nature} \textbf{\bibinfo{volume}{453}},
  \bibinfo{pages}{761} (\bibinfo{year}{2008}{\natexlab{b}}).

\bibitem[{\citenamefont{Rotter et~al.}(2008)\citenamefont{Rotter, Tegel, and
  Johrendt}}]{BaFeAs-rotter}
\bibinfo{author}{\bibfnamefont{M.}~\bibnamefont{Rotter}},
  \bibinfo{author}{\bibfnamefont{M.}~\bibnamefont{Tegel}}, \bibnamefont{and}
  \bibinfo{author}{\bibfnamefont{D.}~\bibnamefont{Johrendt}},
  \bibinfo{journal}{Phys. Rev. Lett.} \textbf{\bibinfo{volume}{101}},
  \bibinfo{pages}{107006} (\bibinfo{year}{2008}).

\bibitem[{\citenamefont{Sefat et~al.}(2008)\citenamefont{Sefat, Jin, McGuire,
  Sales, Singh, and Mandrus}}]{BaFeSe-sefat}
\bibinfo{author}{\bibfnamefont{A.~S.} \bibnamefont{Sefat}},
  \bibinfo{author}{\bibfnamefont{R.}~\bibnamefont{Jin}},
  \bibinfo{author}{\bibfnamefont{M.~A.} \bibnamefont{McGuire}},
  \bibinfo{author}{\bibfnamefont{B.~C.} \bibnamefont{Sales}},
  \bibinfo{author}{\bibfnamefont{D.~J.} \bibnamefont{Singh}}, \bibnamefont{and}
  \bibinfo{author}{\bibfnamefont{D.}~\bibnamefont{Mandrus}},
  \bibinfo{journal}{Phys. Rev. Lett.} \textbf{\bibinfo{volume}{101}},
  \bibinfo{pages}{117004} (\bibinfo{year}{2008}).

\bibitem[{\citenamefont{Sharma et~al.}(2010)\citenamefont{Sharma, Bharathi,
  Chandra, Reddy, Paulraj, Satya, Sastry, Gupta, and Sundar}}]{BaFeRuAs-sharma}
\bibinfo{author}{\bibfnamefont{S.}~\bibnamefont{Sharma}},
  \bibinfo{author}{\bibfnamefont{A.}~\bibnamefont{Bharathi}},
  \bibinfo{author}{\bibfnamefont{S.}~\bibnamefont{Chandra}},
  \bibinfo{author}{\bibfnamefont{V.~R.} \bibnamefont{Reddy}},
  \bibinfo{author}{\bibfnamefont{S.}~\bibnamefont{Paulraj}},
  \bibinfo{author}{\bibfnamefont{A.~T.} \bibnamefont{Satya}},
  \bibinfo{author}{\bibfnamefont{V.~S.} \bibnamefont{Sastry}},
  \bibinfo{author}{\bibfnamefont{A.}~\bibnamefont{Gupta}}, \bibnamefont{and}
  \bibinfo{author}{\bibfnamefont{C.~S.} \bibnamefont{Sundar}},
  \bibinfo{journal}{Phys. Rev. B} \textbf{\bibinfo{volume}{81}},
  \bibinfo{pages}{174512} (\bibinfo{year}{2010}).

\bibitem[{\citenamefont{Tapp et~al.}(2008)\citenamefont{Tapp, Tang, Lv, Sasmal,
  Lorenz, Chu, and Guloy}}]{Tapp-LiFeAs}
\bibinfo{author}{\bibfnamefont{J.~H.} \bibnamefont{Tapp}},
  \bibinfo{author}{\bibfnamefont{Z.}~\bibnamefont{Tang}},
  \bibinfo{author}{\bibfnamefont{B.}~\bibnamefont{Lv}},
  \bibinfo{author}{\bibfnamefont{K.}~\bibnamefont{Sasmal}},
  \bibinfo{author}{\bibfnamefont{B.}~\bibnamefont{Lorenz}},
  \bibinfo{author}{\bibfnamefont{P.~C.~W.} \bibnamefont{Chu}},
  \bibnamefont{and} \bibinfo{author}{\bibfnamefont{A.~M.} \bibnamefont{Guloy}},
  \bibinfo{journal}{Phys. Rev. B} \textbf{\bibinfo{volume}{78}},
  \bibinfo{pages}{060505} (\bibinfo{year}{2008}).

\bibitem[{\citenamefont{Wang et~al.}(2008)\citenamefont{Wang, Liu, Lv, Gao,
  Yang, Yu, Li, and Jin}}]{Wang-LiFeAs}
\bibinfo{author}{\bibfnamefont{X.}~\bibnamefont{Wang}},
  \bibinfo{author}{\bibfnamefont{Q.}~\bibnamefont{Liu}},
  \bibinfo{author}{\bibfnamefont{Y.}~\bibnamefont{Lv}},
  \bibinfo{author}{\bibfnamefont{W.}~\bibnamefont{Gao}},
  \bibinfo{author}{\bibfnamefont{L.}~\bibnamefont{Yang}},
  \bibinfo{author}{\bibfnamefont{R.}~\bibnamefont{Yu}},
  \bibinfo{author}{\bibfnamefont{F.}~\bibnamefont{Li}}, \bibnamefont{and}
  \bibinfo{author}{\bibfnamefont{C.}~\bibnamefont{Jin}},
  \bibinfo{journal}{Solid State Commun.} \textbf{\bibinfo{volume}{148}},
  \bibinfo{pages}{538} (\bibinfo{year}{2008}).

\bibitem[{\citenamefont{Pitcher et~al.}(2008)\citenamefont{Pitcher, Parker,
  Adamson, Herkelrath, Boothroyd, Ibberson, Brunelli, and
  Clarke}}]{Pitcher-LiFeAs}
\bibinfo{author}{\bibfnamefont{M.~J.} \bibnamefont{Pitcher}},
  \bibinfo{author}{\bibfnamefont{D.~R.} \bibnamefont{Parker}},
  \bibinfo{author}{\bibfnamefont{P.}~\bibnamefont{Adamson}},
  \bibinfo{author}{\bibfnamefont{S.~J.} \bibnamefont{Herkelrath}},
  \bibinfo{author}{\bibfnamefont{A.~T.} \bibnamefont{Boothroyd}},
  \bibinfo{author}{\bibfnamefont{R.~M.} \bibnamefont{Ibberson}},
  \bibinfo{author}{\bibfnamefont{M.}~\bibnamefont{Brunelli}}, \bibnamefont{and}
  \bibinfo{author}{\bibfnamefont{S.~J.} \bibnamefont{Clarke}},
  \bibinfo{journal}{Chem. Commun.} pp. \bibinfo{pages}{5918--5920}
  (\bibinfo{year}{2008}).

\bibitem[{\citenamefont{Hsu et~al.}(2008)\citenamefont{Hsu, Luo, Yeh, Chen,
  Huang, Wu, Lee, Huang, Chu, Yan et~al.}}]{FeSe-Hsu}
\bibinfo{author}{\bibfnamefont{F.-C.} \bibnamefont{Hsu}},
  \bibinfo{author}{\bibfnamefont{J.-Y.} \bibnamefont{Luo}},
  \bibinfo{author}{\bibfnamefont{K.-W.} \bibnamefont{Yeh}},
  \bibinfo{author}{\bibfnamefont{T.-K.} \bibnamefont{Chen}},
  \bibinfo{author}{\bibfnamefont{T.-W.} \bibnamefont{Huang}},
  \bibinfo{author}{\bibfnamefont{P.~M.} \bibnamefont{Wu}},
  \bibinfo{author}{\bibfnamefont{Y.-C.} \bibnamefont{Lee}},
  \bibinfo{author}{\bibfnamefont{Y.-L.} \bibnamefont{Huang}},
  \bibinfo{author}{\bibfnamefont{Y.-Y.} \bibnamefont{Chu}},
  \bibinfo{author}{\bibfnamefont{D.-C.} \bibnamefont{Yan}},
  \bibnamefont{et~al.}, \bibinfo{journal}{Proc. Natl. Acad. Sci. U.S.A.}
  \textbf{\bibinfo{volume}{105}}, \bibinfo{pages}{14262}
  (\bibinfo{year}{2008}).

\bibitem[{\citenamefont{Margadonna et~al.}(2009)\citenamefont{Margadonna,
  Takabayashi, Ohishi, Mizuguchi, Takano, Kagayama, Nakagawa, Takata, and
  Prassides}}]{FeSe-underpressure-2009}
\bibinfo{author}{\bibfnamefont{S.}~\bibnamefont{Margadonna}},
  \bibinfo{author}{\bibfnamefont{Y.}~\bibnamefont{Takabayashi}},
  \bibinfo{author}{\bibfnamefont{Y.}~\bibnamefont{Ohishi}},
  \bibinfo{author}{\bibfnamefont{Y.}~\bibnamefont{Mizuguchi}},
  \bibinfo{author}{\bibfnamefont{Y.}~\bibnamefont{Takano}},
  \bibinfo{author}{\bibfnamefont{T.}~\bibnamefont{Kagayama}},
  \bibinfo{author}{\bibfnamefont{T.}~\bibnamefont{Nakagawa}},
  \bibinfo{author}{\bibfnamefont{M.}~\bibnamefont{Takata}}, \bibnamefont{and}
  \bibinfo{author}{\bibfnamefont{K.}~\bibnamefont{Prassides}},
  \bibinfo{journal}{Phys. Rev. B} \textbf{\bibinfo{volume}{80}},
  \bibinfo{pages}{064506} (\bibinfo{year}{2009}).

\bibitem[{\citenamefont{Yeh et~al.}(2008)\citenamefont{Yeh, Huang, Huang, Chen,
  Hsu, Wu, Lee, Chu, Chen, Luo et~al.}}]{FeSeTe-yeh}
\bibinfo{author}{\bibfnamefont{K.-W.} \bibnamefont{Yeh}},
  \bibinfo{author}{\bibfnamefont{T.-W.} \bibnamefont{Huang}},
  \bibinfo{author}{\bibfnamefont{Y.-l.} \bibnamefont{Huang}},
  \bibinfo{author}{\bibfnamefont{T.-K.} \bibnamefont{Chen}},
  \bibinfo{author}{\bibfnamefont{F.}~\bibnamefont{Hsu}},
  \bibinfo{author}{\bibfnamefont{P.~M.} \bibnamefont{Wu}},
  \bibinfo{author}{\bibfnamefont{Y.~C.} \bibnamefont{Lee}},
  \bibinfo{author}{\bibfnamefont{Y.-Y.} \bibnamefont{Chu}},
  \bibinfo{author}{\bibfnamefont{C.-L.} \bibnamefont{Chen}},
  \bibinfo{author}{\bibfnamefont{J.-Y.} \bibnamefont{Luo}},
  \bibnamefont{et~al.}, \bibinfo{journal}{Eur. Phys. Lett.}
  \textbf{\bibinfo{volume}{84}}, \bibinfo{pages}{37002} (\bibinfo{year}{2008}).

\bibitem[{\citenamefont{de' Medici et~al.}(2014)\citenamefont{de' Medici,
  Giovannetti, and Capone}}]{Medici-selective-Mott}
\bibinfo{author}{\bibfnamefont{L.}~\bibnamefont{de' Medici}},
  \bibinfo{author}{\bibfnamefont{G.}~\bibnamefont{Giovannetti}},
  \bibnamefont{and} \bibinfo{author}{\bibfnamefont{M.}~\bibnamefont{Capone}},
  \bibinfo{journal}{Phys. Rev. Lett.} \textbf{\bibinfo{volume}{112}},
  \bibinfo{pages}{177001} (\bibinfo{year}{2014}).

\bibitem[{\citenamefont{Misawa et~al.}(2012)\citenamefont{Misawa, Nakamura, and
  Imada}}]{Imada-correlation-pnictides}
\bibinfo{author}{\bibfnamefont{T.}~\bibnamefont{Misawa}},
  \bibinfo{author}{\bibfnamefont{K.}~\bibnamefont{Nakamura}}, \bibnamefont{and}
  \bibinfo{author}{\bibfnamefont{M.}~\bibnamefont{Imada}},
  \bibinfo{journal}{Phys. Rev. Lett.} \textbf{\bibinfo{volume}{108}},
  \bibinfo{pages}{177007} (\bibinfo{year}{2012}).

\bibitem[{\citenamefont{Yaresko et~al.}(2009)\citenamefont{Yaresko, Liu,
  Antonov, and Andersen}}]{Yaresko-magnetic-FS}
\bibinfo{author}{\bibfnamefont{A.~N.} \bibnamefont{Yaresko}},
  \bibinfo{author}{\bibfnamefont{G.-Q.} \bibnamefont{Liu}},
  \bibinfo{author}{\bibfnamefont{V.~N.} \bibnamefont{Antonov}},
  \bibnamefont{and} \bibinfo{author}{\bibfnamefont{O.~K.}
  \bibnamefont{Andersen}}, \bibinfo{journal}{Phys. Rev. B}
  \textbf{\bibinfo{volume}{79}}, \bibinfo{pages}{144421}
  (\bibinfo{year}{2009}).

\bibitem[{\citenamefont{Mazin et~al.}(2008)\citenamefont{Mazin, Johannes,
  Boeri, Koepernik, and Singh}}]{pnictides-mazin}
\bibinfo{author}{\bibfnamefont{I.~I.} \bibnamefont{Mazin}},
  \bibinfo{author}{\bibfnamefont{M.~D.} \bibnamefont{Johannes}},
  \bibinfo{author}{\bibfnamefont{L.}~\bibnamefont{Boeri}},
  \bibinfo{author}{\bibfnamefont{K.}~\bibnamefont{Koepernik}},
  \bibnamefont{and} \bibinfo{author}{\bibfnamefont{D.~J.} \bibnamefont{Singh}},
  \bibinfo{journal}{Phys. Rev. B} \textbf{\bibinfo{volume}{78}},
  \bibinfo{pages}{085104} (\bibinfo{year}{2008}).

\bibitem[{\citenamefont{Subedi et~al.}(2008)\citenamefont{Subedi, Zhang, Singh,
  and Du}}]{FeSe-alaska}
\bibinfo{author}{\bibfnamefont{A.}~\bibnamefont{Subedi}},
  \bibinfo{author}{\bibfnamefont{L.}~\bibnamefont{Zhang}},
  \bibinfo{author}{\bibfnamefont{D.~J.} \bibnamefont{Singh}}, \bibnamefont{and}
  \bibinfo{author}{\bibfnamefont{M.~H.} \bibnamefont{Du}},
  \bibinfo{journal}{Phys. Rev. B} \textbf{\bibinfo{volume}{78}},
  \bibinfo{pages}{134514} (\bibinfo{year}{2008}).

\bibitem[{\citenamefont{Ishida et~al.}(2009)\citenamefont{Ishida, Nakai, and
  Hosono}}]{ishida-2009}
\bibinfo{author}{\bibfnamefont{K.}~\bibnamefont{Ishida}},
  \bibinfo{author}{\bibfnamefont{Y.}~\bibnamefont{Nakai}}, \bibnamefont{and}
  \bibinfo{author}{\bibfnamefont{H.}~\bibnamefont{Hosono}},
  \bibinfo{journal}{J. Phys. Soc. Jpn.} \textbf{\bibinfo{volume}{78}},
  \bibinfo{pages}{062001} (\bibinfo{year}{2009}).

\bibitem[{\citenamefont{Qureshi et~al.}(2010)\citenamefont{Qureshi, Drees,
  Werner, Wurmehl, Hess, Klingeler, B\"uchner, Fernandez-Diaz, and
  Braden}}]{LaOFeAs-qureshi-2010}
\bibinfo{author}{\bibfnamefont{N.}~\bibnamefont{Qureshi}},
  \bibinfo{author}{\bibfnamefont{Y.}~\bibnamefont{Drees}},
  \bibinfo{author}{\bibfnamefont{J.}~\bibnamefont{Werner}},
  \bibinfo{author}{\bibfnamefont{S.}~\bibnamefont{Wurmehl}},
  \bibinfo{author}{\bibfnamefont{C.}~\bibnamefont{Hess}},
  \bibinfo{author}{\bibfnamefont{R.}~\bibnamefont{Klingeler}},
  \bibinfo{author}{\bibfnamefont{B.}~\bibnamefont{B\"uchner}},
  \bibinfo{author}{\bibfnamefont{M.~T.} \bibnamefont{Fernandez-Diaz}},
  \bibnamefont{and} \bibinfo{author}{\bibfnamefont{M.}~\bibnamefont{Braden}},
  \bibinfo{journal}{Phys. Rev. B} \textbf{\bibinfo{volume}{82}},
  \bibinfo{pages}{184521} (\bibinfo{year}{2010}).

\bibitem[{\citenamefont{Sawatzky et~al.}(2009)\citenamefont{Sawatzky, Elfimov,
  van~den Brink, and Zaanen}}]{LaOFeAs-Sawatzky}
\bibinfo{author}{\bibfnamefont{G.~A.} \bibnamefont{Sawatzky}},
  \bibinfo{author}{\bibfnamefont{I.~S.} \bibnamefont{Elfimov}},
  \bibinfo{author}{\bibfnamefont{J.}~\bibnamefont{van~den Brink}},
  \bibnamefont{and} \bibinfo{author}{\bibfnamefont{J.}~\bibnamefont{Zaanen}},
  \bibinfo{journal}{Eur. Phys. Lett.} \textbf{\bibinfo{volume}{86}},
  \bibinfo{pages}{17006} (\bibinfo{year}{2009}).

\bibitem[{\citenamefont{Huang et~al.}(2008)\citenamefont{Huang, Qiu, Bao,
  Green, Lynn, Gasparovic, Wu, Wu, and Chen}}]{BaFeAs-huang}
\bibinfo{author}{\bibfnamefont{Q.}~\bibnamefont{Huang}},
  \bibinfo{author}{\bibfnamefont{Y.}~\bibnamefont{Qiu}},
  \bibinfo{author}{\bibfnamefont{W.}~\bibnamefont{Bao}},
  \bibinfo{author}{\bibfnamefont{M.~A.} \bibnamefont{Green}},
  \bibinfo{author}{\bibfnamefont{J.~W.} \bibnamefont{Lynn}},
  \bibinfo{author}{\bibfnamefont{Y.~C.} \bibnamefont{Gasparovic}},
  \bibinfo{author}{\bibfnamefont{T.}~\bibnamefont{Wu}},
  \bibinfo{author}{\bibfnamefont{G.}~\bibnamefont{Wu}}, \bibnamefont{and}
  \bibinfo{author}{\bibfnamefont{X.~H.} \bibnamefont{Chen}},
  \bibinfo{journal}{Phys. Rev. Lett.} \textbf{\bibinfo{volume}{101}},
  \bibinfo{pages}{257003} (\bibinfo{year}{2008}).

\bibitem[{\citenamefont{Li et~al.}(2009)\citenamefont{Li, de~la Cruz, Huang,
  Chen, Lynn, Hu, Huang, Hsu, Yeh, Wu et~al.}}]{FeTe-mu-li}
\bibinfo{author}{\bibfnamefont{S.}~\bibnamefont{Li}},
  \bibinfo{author}{\bibfnamefont{C.}~\bibnamefont{de~la Cruz}},
  \bibinfo{author}{\bibfnamefont{Q.}~\bibnamefont{Huang}},
  \bibinfo{author}{\bibfnamefont{Y.}~\bibnamefont{Chen}},
  \bibinfo{author}{\bibfnamefont{J.~W.} \bibnamefont{Lynn}},
  \bibinfo{author}{\bibfnamefont{J.}~\bibnamefont{Hu}},
  \bibinfo{author}{\bibfnamefont{Y.-L.} \bibnamefont{Huang}},
  \bibinfo{author}{\bibfnamefont{F.-C.} \bibnamefont{Hsu}},
  \bibinfo{author}{\bibfnamefont{K.-W.} \bibnamefont{Yeh}},
  \bibinfo{author}{\bibfnamefont{M.-K.} \bibnamefont{Wu}},
  \bibnamefont{et~al.}, \bibinfo{journal}{Phys. Rev. B}
  \textbf{\bibinfo{volume}{79}}, \bibinfo{pages}{054503}
  (\bibinfo{year}{2009}).

\bibitem[{\citenamefont{Hansmann et~al.}(2010)\citenamefont{Hansmann, Arita,
  Toschi, Sakai, Sangiovanni, and Held}}]{LaOFeAs-hansmann-2010}
\bibinfo{author}{\bibfnamefont{P.}~\bibnamefont{Hansmann}},
  \bibinfo{author}{\bibfnamefont{R.}~\bibnamefont{Arita}},
  \bibinfo{author}{\bibfnamefont{A.}~\bibnamefont{Toschi}},
  \bibinfo{author}{\bibfnamefont{S.}~\bibnamefont{Sakai}},
  \bibinfo{author}{\bibfnamefont{G.}~\bibnamefont{Sangiovanni}},
  \bibnamefont{and} \bibinfo{author}{\bibfnamefont{K.}~\bibnamefont{Held}},
  \bibinfo{journal}{Phys. Rev. Lett.} \textbf{\bibinfo{volume}{104}},
  \bibinfo{pages}{197002} (\bibinfo{year}{2010}).

\bibitem[{\citenamefont{Toschi et~al.}(2012)\citenamefont{Toschi, Arita,
  Hansmann, Sangiovanni, and Held}}]{LaOFeAs-hansmann-2012}
\bibinfo{author}{\bibfnamefont{A.}~\bibnamefont{Toschi}},
  \bibinfo{author}{\bibfnamefont{R.}~\bibnamefont{Arita}},
  \bibinfo{author}{\bibfnamefont{P.}~\bibnamefont{Hansmann}},
  \bibinfo{author}{\bibfnamefont{G.}~\bibnamefont{Sangiovanni}},
  \bibnamefont{and} \bibinfo{author}{\bibfnamefont{K.}~\bibnamefont{Held}},
  \bibinfo{journal}{Phys. Rev. B} \textbf{\bibinfo{volume}{86}},
  \bibinfo{pages}{064411} (\bibinfo{year}{2012}).

\bibitem[{\citenamefont{{Hansmann} et~al.}(2015)\citenamefont{{Hansmann},
  {Ayral}, {Tejeda}, and {Biermann}}}]{Hansmann_SciRep}
\bibinfo{author}{\bibfnamefont{P.}~\bibnamefont{{Hansmann}}},
  \bibinfo{author}{\bibfnamefont{T.}~\bibnamefont{{Ayral}}},
  \bibinfo{author}{\bibfnamefont{A.}~\bibnamefont{{Tejeda}}}, \bibnamefont{and}
  \bibinfo{author}{\bibfnamefont{S.}~\bibnamefont{{Biermann}}},
  \bibinfo{journal}{arXiv: 1511.05004}  (\bibinfo{year}{2015}).

\bibitem[{\citenamefont{Georges et~al.}(1996)\citenamefont{Georges, Kotliar,
  Krauth, and Rozenberg}}]{RevDMFT_AG}
\bibinfo{author}{\bibfnamefont{A.}~\bibnamefont{Georges}},
  \bibinfo{author}{\bibfnamefont{G.}~\bibnamefont{Kotliar}},
  \bibinfo{author}{\bibfnamefont{W.}~\bibnamefont{Krauth}}, \bibnamefont{and}
  \bibinfo{author}{\bibfnamefont{M.~J.} \bibnamefont{Rozenberg}},
  \bibinfo{journal}{Rev. Mod. Phys.} \textbf{\bibinfo{volume}{68}},
  \bibinfo{pages}{13} (\bibinfo{year}{1996}).

\bibitem[{\citenamefont{Kotliar et~al.}(2006)\citenamefont{Kotliar, Savrasov,
  Haule, Oudovenko, Parcollet, and Marianetti}}]{kotliar-review-DMFT}
\bibinfo{author}{\bibfnamefont{G.}~\bibnamefont{Kotliar}},
  \bibinfo{author}{\bibfnamefont{S.~Y.} \bibnamefont{Savrasov}},
  \bibinfo{author}{\bibfnamefont{K.}~\bibnamefont{Haule}},
  \bibinfo{author}{\bibfnamefont{V.~S.} \bibnamefont{Oudovenko}},
  \bibinfo{author}{\bibfnamefont{O.}~\bibnamefont{Parcollet}},
  \bibnamefont{and} \bibinfo{author}{\bibfnamefont{C.~A.}
  \bibnamefont{Marianetti}}, \bibinfo{journal}{Rev. Mod. Phys.}
  \textbf{\bibinfo{volume}{78}}, \bibinfo{pages}{865} (\bibinfo{year}{2006}).

\bibitem[{\citenamefont{Anisimov
  et~al.}(1997{\natexlab{a}})\citenamefont{Anisimov, Poteryaev, Korotin,
  Anokhin, and Kotliar}}]{LDA+DMFT-anisimov-1997}
\bibinfo{author}{\bibfnamefont{V.~I.} \bibnamefont{Anisimov}},
  \bibinfo{author}{\bibfnamefont{A.}~\bibnamefont{Poteryaev}},
  \bibinfo{author}{\bibfnamefont{M.}~\bibnamefont{Korotin}},
  \bibinfo{author}{\bibfnamefont{A.}~\bibnamefont{Anokhin}}, \bibnamefont{and}
  \bibinfo{author}{\bibfnamefont{G.}~\bibnamefont{Kotliar}},
  \bibinfo{journal}{J. Phys.: Condens. Matter} \textbf{\bibinfo{volume}{9}},
  \bibinfo{pages}{943} (\bibinfo{year}{1997}{\natexlab{a}}).

\bibitem[{\citenamefont{Lichtenstein and Katsnelson}(1998)}]{LDA+DMFT-licht}
\bibinfo{author}{\bibfnamefont{A.~I.} \bibnamefont{Lichtenstein}}
  \bibnamefont{and} \bibinfo{author}{\bibfnamefont{M.~I.}
  \bibnamefont{Katsnelson}}, \bibinfo{journal}{Phys. Rev. B}
  \textbf{\bibinfo{volume}{57}}, \bibinfo{pages}{6884} (\bibinfo{year}{1998}).

\bibitem[{\citenamefont{Anisimov et~al.}(2009)\citenamefont{Anisimov, Korotin,
  Korotin, Kozhevnikov, Kunes, O., Skornyakov, and
  Streltsov}}]{LaOFeAs-anisimov-2009}
\bibinfo{author}{\bibfnamefont{V.~I.} \bibnamefont{Anisimov}},
  \bibinfo{author}{\bibfnamefont{D.}~\bibnamefont{Korotin}},
  \bibinfo{author}{\bibfnamefont{M.}~\bibnamefont{Korotin}},
  \bibinfo{author}{\bibfnamefont{A.~V.} \bibnamefont{Kozhevnikov}},
  \bibinfo{author}{\bibfnamefont{J.}~\bibnamefont{Kunes}},
  \bibinfo{author}{\bibfnamefont{S.~A.} \bibnamefont{O.}},
  \bibinfo{author}{\bibfnamefont{S.~L.} \bibnamefont{Skornyakov}},
  \bibnamefont{and} \bibinfo{author}{\bibfnamefont{S.~V.}
  \bibnamefont{Streltsov}}, \bibinfo{journal}{J. Phys. Condens. Matter}
  \textbf{\bibinfo{volume}{21}}, \bibinfo{pages}{075602}
  (\bibinfo{year}{2009}).

\bibitem[{\citenamefont{Shorikov et~al.}(2009)\citenamefont{Shorikov, Korotin,
  Streltsov, Skornyakov, Korotin, and Anisimov}}]{LaOFeAs-shorikov}
\bibinfo{author}{\bibfnamefont{A.~O.} \bibnamefont{Shorikov}},
  \bibinfo{author}{\bibfnamefont{M.~A.} \bibnamefont{Korotin}},
  \bibinfo{author}{\bibfnamefont{S.~V.} \bibnamefont{Streltsov}},
  \bibinfo{author}{\bibfnamefont{S.~L.} \bibnamefont{Skornyakov}},
  \bibinfo{author}{\bibfnamefont{D.~M.} \bibnamefont{Korotin}},
  \bibnamefont{and} \bibinfo{author}{\bibfnamefont{V.~I.}
  \bibnamefont{Anisimov}}, \bibinfo{journal}{J. Exp. Theor. Phys.}
  \textbf{\bibinfo{volume}{108}}, \bibinfo{pages}{121} (\bibinfo{year}{2009}).

\bibitem[{\citenamefont{Anisimov et~al.}(2008)\citenamefont{Anisimov, Korotin,
  Streltsov, Kozhevnikov, Kunes, O., and Korotin}}]{LaOFeAs-anisimov-2008}
\bibinfo{author}{\bibfnamefont{V.~I.} \bibnamefont{Anisimov}},
  \bibinfo{author}{\bibfnamefont{D.~M.} \bibnamefont{Korotin}},
  \bibinfo{author}{\bibfnamefont{S.~V.} \bibnamefont{Streltsov}},
  \bibinfo{author}{\bibfnamefont{A.~V.} \bibnamefont{Kozhevnikov}},
  \bibinfo{author}{\bibfnamefont{J.}~\bibnamefont{Kunes}},
  \bibinfo{author}{\bibfnamefont{S.~A.} \bibnamefont{O.}}, \bibnamefont{and}
  \bibinfo{author}{\bibfnamefont{M.~A.} \bibnamefont{Korotin}},
  \bibinfo{journal}{JETP Lett.} \textbf{\bibinfo{volume}{88}},
  \bibinfo{pages}{729} (\bibinfo{year}{2008}).

\bibitem[{\citenamefont{Aichhorn et~al.}(2009)\citenamefont{Aichhorn,
  Pourovskii, Vildosola, Ferrero, Parcollet, Miyake, Georges, and
  Biermann}}]{cRPA-DMFT-LaOFeAs-markus}
\bibinfo{author}{\bibfnamefont{M.}~\bibnamefont{Aichhorn}},
  \bibinfo{author}{\bibfnamefont{L.}~\bibnamefont{Pourovskii}},
  \bibinfo{author}{\bibfnamefont{V.}~\bibnamefont{Vildosola}},
  \bibinfo{author}{\bibfnamefont{M.}~\bibnamefont{Ferrero}},
  \bibinfo{author}{\bibfnamefont{O.}~\bibnamefont{Parcollet}},
  \bibinfo{author}{\bibfnamefont{T.}~\bibnamefont{Miyake}},
  \bibinfo{author}{\bibfnamefont{A.}~\bibnamefont{Georges}}, \bibnamefont{and}
  \bibinfo{author}{\bibfnamefont{S.}~\bibnamefont{Biermann}},
  \bibinfo{journal}{Phys. Rev. B} \textbf{\bibinfo{volume}{80}},
  \bibinfo{pages}{085101} (\bibinfo{year}{2009}).

\bibitem[{\citenamefont{Haule et~al.}(2008)\citenamefont{Haule, Shim, and
  Kotliar}}]{LaOFeAs-haule-2008}
\bibinfo{author}{\bibfnamefont{K.}~\bibnamefont{Haule}},
  \bibinfo{author}{\bibfnamefont{J.~H.} \bibnamefont{Shim}}, \bibnamefont{and}
  \bibinfo{author}{\bibfnamefont{G.}~\bibnamefont{Kotliar}},
  \bibinfo{journal}{Phys. Rev. Lett.} \textbf{\bibinfo{volume}{100}},
  \bibinfo{pages}{226402} (\bibinfo{year}{2008}).

\bibitem[{\citenamefont{Haule and Kotliar}(2009)}]{LaOFeAs-kotliar-2009}
\bibinfo{author}{\bibfnamefont{K.}~\bibnamefont{Haule}} \bibnamefont{and}
  \bibinfo{author}{\bibfnamefont{G.}~\bibnamefont{Kotliar}},
  \bibinfo{journal}{New J. Phys.} \textbf{\bibinfo{volume}{11}},
  \bibinfo{pages}{025021} (\bibinfo{year}{2009}).

\bibitem[{\citenamefont{Liebsch and Ishida}(2010)}]{liebsch-FeSe}
\bibinfo{author}{\bibfnamefont{A.}~\bibnamefont{Liebsch}} \bibnamefont{and}
  \bibinfo{author}{\bibfnamefont{H.}~\bibnamefont{Ishida}},
  \bibinfo{journal}{Phys. Rev. B} \textbf{\bibinfo{volume}{82}},
  \bibinfo{pages}{155106} (\bibinfo{year}{2010}).

\bibitem[{\citenamefont{Aichhorn et~al.}(2010)\citenamefont{Aichhorn, Biermann,
  Miyake, Georges, and Imada}}]{cRPA-DMFT-FeSe-markus}
\bibinfo{author}{\bibfnamefont{M.}~\bibnamefont{Aichhorn}},
  \bibinfo{author}{\bibfnamefont{S.}~\bibnamefont{Biermann}},
  \bibinfo{author}{\bibfnamefont{T.}~\bibnamefont{Miyake}},
  \bibinfo{author}{\bibfnamefont{A.}~\bibnamefont{Georges}}, \bibnamefont{and}
  \bibinfo{author}{\bibfnamefont{M.}~\bibnamefont{Imada}},
  \bibinfo{journal}{Phys. Rev. B} \textbf{\bibinfo{volume}{82}},
  \bibinfo{pages}{064504} (\bibinfo{year}{2010}).

\bibitem[{\citenamefont{Haule et~al.}(2010)\citenamefont{Haule, Yee, and
  Kim}}]{sc-LDA+DMFT-haule}
\bibinfo{author}{\bibfnamefont{K.}~\bibnamefont{Haule}},
  \bibinfo{author}{\bibfnamefont{C.-H.} \bibnamefont{Yee}}, \bibnamefont{and}
  \bibinfo{author}{\bibfnamefont{K.}~\bibnamefont{Kim}},
  \bibinfo{journal}{Phys. Rev. B} \textbf{\bibinfo{volume}{81}},
  \bibinfo{pages}{195107} (\bibinfo{year}{2010}).

\bibitem[{\citenamefont{Aryasetiawan et~al.}(2004)\citenamefont{Aryasetiawan,
  Imada, Georges, Kotliar, Biermann, and Lichtenstein}}]{cRPA-ferdi-2004}
\bibinfo{author}{\bibfnamefont{F.}~\bibnamefont{Aryasetiawan}},
  \bibinfo{author}{\bibfnamefont{M.}~\bibnamefont{Imada}},
  \bibinfo{author}{\bibfnamefont{A.}~\bibnamefont{Georges}},
  \bibinfo{author}{\bibfnamefont{G.}~\bibnamefont{Kotliar}},
  \bibinfo{author}{\bibfnamefont{S.}~\bibnamefont{Biermann}}, \bibnamefont{and}
  \bibinfo{author}{\bibfnamefont{A.~I.} \bibnamefont{Lichtenstein}},
  \bibinfo{journal}{Phys. Rev. B} \textbf{\bibinfo{volume}{70}},
  \bibinfo{pages}{195104} (\bibinfo{year}{2004}).

\bibitem[{\citenamefont{Miyake et~al.}(2008)\citenamefont{Miyake, Pourovskii,
  Vildosola, Biermann, and Georges}}]{cRPA-LaOFeAs-miyake}
\bibinfo{author}{\bibfnamefont{T.}~\bibnamefont{Miyake}},
  \bibinfo{author}{\bibfnamefont{L.}~\bibnamefont{Pourovskii}},
  \bibinfo{author}{\bibfnamefont{V.}~\bibnamefont{Vildosola}},
  \bibinfo{author}{\bibfnamefont{S.}~\bibnamefont{Biermann}}, \bibnamefont{and}
  \bibinfo{author}{\bibfnamefont{A.}~\bibnamefont{Georges}},
  \bibinfo{journal}{J. Phys. Soc. Jpn. : Supplement C}
  \textbf{\bibinfo{volume}{77}}, \bibinfo{pages}{99} (\bibinfo{year}{2008}).

\bibitem[{\citenamefont{Miyake et~al.}(2010)\citenamefont{Miyake, Nakamura,
  Arita, and Imada}}]{cRPA-pnictides-takashi}
\bibinfo{author}{\bibfnamefont{T.}~\bibnamefont{Miyake}},
  \bibinfo{author}{\bibfnamefont{K.}~\bibnamefont{Nakamura}},
  \bibinfo{author}{\bibfnamefont{R.}~\bibnamefont{Arita}}, \bibnamefont{and}
  \bibinfo{author}{\bibfnamefont{M.}~\bibnamefont{Imada}}, \bibinfo{journal}{J.
  Phys. Soc. Jpn.} \textbf{\bibinfo{volume}{79}}, \bibinfo{pages}{044705}
  (\bibinfo{year}{2010}).

\bibitem[{\citenamefont{Casula et~al.}(2012{\natexlab{a}})\citenamefont{Casula,
  Werner, Vaugier, Aryasetiawan, Miyake, Millis, and
  Biermann}}]{udyneff-michele}
\bibinfo{author}{\bibfnamefont{M.}~\bibnamefont{Casula}},
  \bibinfo{author}{\bibfnamefont{P.}~\bibnamefont{Werner}},
  \bibinfo{author}{\bibfnamefont{L.}~\bibnamefont{Vaugier}},
  \bibinfo{author}{\bibfnamefont{F.}~\bibnamefont{Aryasetiawan}},
  \bibinfo{author}{\bibfnamefont{T.}~\bibnamefont{Miyake}},
  \bibinfo{author}{\bibfnamefont{A.~J.} \bibnamefont{Millis}},
  \bibnamefont{and} \bibinfo{author}{\bibfnamefont{S.}~\bibnamefont{Biermann}},
  \bibinfo{journal}{Phys. Rev. Lett.} \textbf{\bibinfo{volume}{109}},
  \bibinfo{pages}{126408} (\bibinfo{year}{2012}{\natexlab{a}}).

\bibitem[{\citenamefont{Nakamura et~al.}(2008)\citenamefont{Nakamura, Arita,
  and Imada}}]{cRPA-LaOFeAs-nakamura}
\bibinfo{author}{\bibfnamefont{K.}~\bibnamefont{Nakamura}},
  \bibinfo{author}{\bibfnamefont{R.}~\bibnamefont{Arita}}, \bibnamefont{and}
  \bibinfo{author}{\bibfnamefont{M.}~\bibnamefont{Imada}}, \bibinfo{journal}{J.
  Phys. Soc. Jpn.} \textbf{\bibinfo{volume}{77}}, \bibinfo{pages}{093711}
  (\bibinfo{year}{2008}).

\bibitem[{\citenamefont{Vaugier}(2011)}]{loig-tesis}
\bibinfo{author}{\bibfnamefont{L.}~\bibnamefont{Vaugier}}, Ph.D. thesis,
  \bibinfo{school}{Ecole Polytechnique, France} (\bibinfo{year}{2011}).

\bibitem[{\citenamefont{Ma et~al.}(2014)\citenamefont{Ma, van Roekeghem,
  Richard, Liu, Miao, Zeng, Xu, Shi, Cao, He et~al.}}]{Ambroise-Ba2Ti2Fe2As4O}
\bibinfo{author}{\bibfnamefont{J.-Z.} \bibnamefont{Ma}},
  \bibinfo{author}{\bibfnamefont{A.}~\bibnamefont{van Roekeghem}},
  \bibinfo{author}{\bibfnamefont{P.}~\bibnamefont{Richard}},
  \bibinfo{author}{\bibfnamefont{Z.-H.} \bibnamefont{Liu}},
  \bibinfo{author}{\bibfnamefont{H.}~\bibnamefont{Miao}},
  \bibinfo{author}{\bibfnamefont{L.-K.} \bibnamefont{Zeng}},
  \bibinfo{author}{\bibfnamefont{N.}~\bibnamefont{Xu}},
  \bibinfo{author}{\bibfnamefont{M.}~\bibnamefont{Shi}},
  \bibinfo{author}{\bibfnamefont{C.}~\bibnamefont{Cao}},
  \bibinfo{author}{\bibfnamefont{J.-B.} \bibnamefont{He}},
  \bibnamefont{et~al.}, \bibinfo{journal}{Phys. Rev. Lett.}
  \textbf{\bibinfo{volume}{113}}, \bibinfo{pages}{266407}
  (\bibinfo{year}{2014}).

\bibitem[{\citenamefont{van Roekeghem et~al.}(2014)\citenamefont{van Roekeghem,
  Ayral, Tomczak, Casula, Xu, Ding, Ferrero, Parcollet, Jiang, and
  Biermann}}]{Ambroise-BaCo2As2}
\bibinfo{author}{\bibfnamefont{A.}~\bibnamefont{van Roekeghem}},
  \bibinfo{author}{\bibfnamefont{T.}~\bibnamefont{Ayral}},
  \bibinfo{author}{\bibfnamefont{J.~M.} \bibnamefont{Tomczak}},
  \bibinfo{author}{\bibfnamefont{M.}~\bibnamefont{Casula}},
  \bibinfo{author}{\bibfnamefont{N.}~\bibnamefont{Xu}},
  \bibinfo{author}{\bibfnamefont{H.}~\bibnamefont{Ding}},
  \bibinfo{author}{\bibfnamefont{M.}~\bibnamefont{Ferrero}},
  \bibinfo{author}{\bibfnamefont{O.}~\bibnamefont{Parcollet}},
  \bibinfo{author}{\bibfnamefont{H.}~\bibnamefont{Jiang}}, \bibnamefont{and}
  \bibinfo{author}{\bibfnamefont{S.}~\bibnamefont{Biermann}},
  \bibinfo{journal}{Phys. Rev. Lett.} \textbf{\bibinfo{volume}{113}},
  \bibinfo{pages}{266403} (\bibinfo{year}{2014}).

\bibitem[{\citenamefont{van Roekeghem et~al.}(2015)\citenamefont{van Roekeghem,
  Richard, Shi, Wu, Zeng, Saparov, Ohtsubo, Qian, Safa-Sefat, Biermann
  et~al.}}]{Ambroise-CaFe2As2}
\bibinfo{author}{\bibfnamefont{A.}~\bibnamefont{van Roekeghem}},
  \bibinfo{author}{\bibfnamefont{P.}~\bibnamefont{Richard}},
  \bibinfo{author}{\bibfnamefont{X.}~\bibnamefont{Shi}},
  \bibinfo{author}{\bibfnamefont{S.-F.} \bibnamefont{Wu}},
  \bibinfo{author}{\bibfnamefont{L.-K.} \bibnamefont{Zeng}},
  \bibinfo{author}{\bibfnamefont{B.~I.} \bibnamefont{Saparov}},
  \bibinfo{author}{\bibfnamefont{Y.}~\bibnamefont{Ohtsubo}},
  \bibinfo{author}{\bibfnamefont{T.}~\bibnamefont{Qian}},
  \bibinfo{author}{\bibfnamefont{A.}~\bibnamefont{Safa-Sefat}},
  \bibinfo{author}{\bibfnamefont{S.}~\bibnamefont{Biermann}},
  \bibnamefont{et~al.}, \bibinfo{journal}{arXiv:1505.00753}
  (\bibinfo{year}{2015}).

\bibitem[{\citenamefont{Xu et~al.}(2013)\citenamefont{Xu, Richard, van
  Roekeghem, Zhang, Miao, Zhang, Qian, Ferrero, Sefat, Biermann
  et~al.}}]{BaCo2As2-Nan}
\bibinfo{author}{\bibfnamefont{N.}~\bibnamefont{Xu}},
  \bibinfo{author}{\bibfnamefont{P.}~\bibnamefont{Richard}},
  \bibinfo{author}{\bibfnamefont{A.}~\bibnamefont{van Roekeghem}},
  \bibinfo{author}{\bibfnamefont{P.}~\bibnamefont{Zhang}},
  \bibinfo{author}{\bibfnamefont{H.}~\bibnamefont{Miao}},
  \bibinfo{author}{\bibfnamefont{W.-L.} \bibnamefont{Zhang}},
  \bibinfo{author}{\bibfnamefont{T.}~\bibnamefont{Qian}},
  \bibinfo{author}{\bibfnamefont{M.}~\bibnamefont{Ferrero}},
  \bibinfo{author}{\bibfnamefont{A.~S.} \bibnamefont{Sefat}},
  \bibinfo{author}{\bibfnamefont{S.}~\bibnamefont{Biermann}},
  \bibnamefont{et~al.}, \bibinfo{journal}{Phys. Rev. X}
  \textbf{\bibinfo{volume}{3}}, \bibinfo{pages}{011006} (\bibinfo{year}{2013}).

\bibitem[{\citenamefont{Razzoli et~al.}(2015)\citenamefont{Razzoli, Matt,
  Kobayashi, Wang, Strocov, van Roekeghem, Biermann, Plumb, Radovic, Schmitt
  et~al.}}]{Razzoli-pnictides-correlations}
\bibinfo{author}{\bibfnamefont{E.}~\bibnamefont{Razzoli}},
  \bibinfo{author}{\bibfnamefont{C.~E.} \bibnamefont{Matt}},
  \bibinfo{author}{\bibfnamefont{M.}~\bibnamefont{Kobayashi}},
  \bibinfo{author}{\bibfnamefont{X.-P.} \bibnamefont{Wang}},
  \bibinfo{author}{\bibfnamefont{V.~N.} \bibnamefont{Strocov}},
  \bibinfo{author}{\bibfnamefont{A.}~\bibnamefont{van Roekeghem}},
  \bibinfo{author}{\bibfnamefont{S.}~\bibnamefont{Biermann}},
  \bibinfo{author}{\bibfnamefont{N.~C.} \bibnamefont{Plumb}},
  \bibinfo{author}{\bibfnamefont{M.}~\bibnamefont{Radovic}},
  \bibinfo{author}{\bibfnamefont{T.}~\bibnamefont{Schmitt}},
  \bibnamefont{et~al.}, \bibinfo{journal}{Phys. Rev. B}
  \textbf{\bibinfo{volume}{91}}, \bibinfo{pages}{214502}
  (\bibinfo{year}{2015}).

\bibitem[{\citenamefont{Werner et~al.}(2012)\citenamefont{Werner, Casula,
  Miyake, Aryasetiawan, Millis, and Biermann}}]{udyn-werner}
\bibinfo{author}{\bibfnamefont{P.}~\bibnamefont{Werner}},
  \bibinfo{author}{\bibfnamefont{M.}~\bibnamefont{Casula}},
  \bibinfo{author}{\bibfnamefont{T.}~\bibnamefont{Miyake}},
  \bibinfo{author}{\bibfnamefont{F.}~\bibnamefont{Aryasetiawan}},
  \bibinfo{author}{\bibfnamefont{A.~J.} \bibnamefont{Millis}},
  \bibnamefont{and} \bibinfo{author}{\bibfnamefont{S.}~\bibnamefont{Biermann}},
  \bibinfo{journal}{Nature Physics} \textbf{\bibinfo{volume}{8}},
  \bibinfo{pages}{331} (\bibinfo{year}{2012}).

\bibitem[{\citenamefont{Yoshida et~al.}(2009)\citenamefont{Yoshida, Wakita,
  Okazaki, Mizuguchi, Tsuda, Takano, Takeya, Hirata, Muro, Okawa
  et~al.}}]{FeSe-PES-Yoshida}
\bibinfo{author}{\bibfnamefont{R.}~\bibnamefont{Yoshida}},
  \bibinfo{author}{\bibfnamefont{T.}~\bibnamefont{Wakita}},
  \bibinfo{author}{\bibfnamefont{H.}~\bibnamefont{Okazaki}},
  \bibinfo{author}{\bibfnamefont{Y.}~\bibnamefont{Mizuguchi}},
  \bibinfo{author}{\bibfnamefont{S.}~\bibnamefont{Tsuda}},
  \bibinfo{author}{\bibfnamefont{Y.}~\bibnamefont{Takano}},
  \bibinfo{author}{\bibfnamefont{H.}~\bibnamefont{Takeya}},
  \bibinfo{author}{\bibfnamefont{K.}~\bibnamefont{Hirata}},
  \bibinfo{author}{\bibfnamefont{T.}~\bibnamefont{Muro}},
  \bibinfo{author}{\bibfnamefont{M.}~\bibnamefont{Okawa}},
  \bibnamefont{et~al.}, \bibinfo{journal}{J. Phys. Soc. Jpn.}
  \textbf{\bibinfo{volume}{78}}, \bibinfo{pages}{034708}
  (\bibinfo{year}{2009}).

\bibitem[{\citenamefont{Tamai et~al.}(2010)\citenamefont{Tamai, Ganin,
  Rozbicki, Bacsa, Meevasana, King, Caffio, Schaub, Margadonna, Prassides
  et~al.}}]{FeSe-tamai}
\bibinfo{author}{\bibfnamefont{A.}~\bibnamefont{Tamai}},
  \bibinfo{author}{\bibfnamefont{A.~Y.} \bibnamefont{Ganin}},
  \bibinfo{author}{\bibfnamefont{E.}~\bibnamefont{Rozbicki}},
  \bibinfo{author}{\bibfnamefont{J.}~\bibnamefont{Bacsa}},
  \bibinfo{author}{\bibfnamefont{W.}~\bibnamefont{Meevasana}},
  \bibinfo{author}{\bibfnamefont{P.~D.~C.} \bibnamefont{King}},
  \bibinfo{author}{\bibfnamefont{M.}~\bibnamefont{Caffio}},
  \bibinfo{author}{\bibfnamefont{R.}~\bibnamefont{Schaub}},
  \bibinfo{author}{\bibfnamefont{S.}~\bibnamefont{Margadonna}},
  \bibinfo{author}{\bibfnamefont{K.}~\bibnamefont{Prassides}},
  \bibnamefont{et~al.}, \bibinfo{journal}{Phys. Rev. Lett.}
  \textbf{\bibinfo{volume}{104}}, \bibinfo{pages}{097002}
  (\bibinfo{year}{2010}).

\bibitem[{\citenamefont{Biermann}(2014)}]{Biermann_JPCM_review}
\bibinfo{author}{\bibfnamefont{S.}~\bibnamefont{Biermann}},
  \bibinfo{journal}{J. Phys. Condens. Matter} \textbf{\bibinfo{volume}{26}},
  \bibinfo{pages}{173202} (\bibinfo{year}{2014}).

\bibitem[{\citenamefont{Casula et~al.}(2012{\natexlab{b}})\citenamefont{Casula,
  Rubtsov, and Biermann}}]{udyn-michele}
\bibinfo{author}{\bibfnamefont{M.}~\bibnamefont{Casula}},
  \bibinfo{author}{\bibfnamefont{A.}~\bibnamefont{Rubtsov}}, \bibnamefont{and}
  \bibinfo{author}{\bibfnamefont{S.}~\bibnamefont{Biermann}},
  \bibinfo{journal}{Phys. Rev. B} \textbf{\bibinfo{volume}{85}},
  \bibinfo{pages}{035115} (\bibinfo{year}{2012}{\natexlab{b}}).

\bibitem[{\citenamefont{van Roekeghem and Biermann}(2014)}]{Ambroise-SrVO3}
\bibinfo{author}{\bibfnamefont{A.}~\bibnamefont{van Roekeghem}}
  \bibnamefont{and} \bibinfo{author}{\bibfnamefont{S.}~\bibnamefont{Biermann}},
  \bibinfo{journal}{EPL} \textbf{\bibinfo{volume}{108}}, \bibinfo{pages}{75003}
  (\bibinfo{year}{2014}).

\bibitem[{\citenamefont{Biermann and van Roekeghem}(2015)}]{Silke_DD_polarons}
\bibinfo{author}{\bibfnamefont{S.}~\bibnamefont{Biermann}} \bibnamefont{and}
  \bibinfo{author}{\bibfnamefont{A.}~\bibnamefont{van Roekeghem}},
  \bibinfo{journal}{arXiv:1512.08499, J. El. Sp.}  (\bibinfo{year}{2015}).

\bibitem[{\citenamefont{Vaugier et~al.}(2012)\citenamefont{Vaugier, Jiang, and
  Biermann}}]{TMO-vaugier}
\bibinfo{author}{\bibfnamefont{L.}~\bibnamefont{Vaugier}},
  \bibinfo{author}{\bibfnamefont{H.}~\bibnamefont{Jiang}}, \bibnamefont{and}
  \bibinfo{author}{\bibfnamefont{S.}~\bibnamefont{Biermann}},
  \bibinfo{journal}{Phys. Rev. B} \textbf{\bibinfo{volume}{86}},
  \bibinfo{pages}{165105} (\bibinfo{year}{2012}).

\bibitem[{\citenamefont{Blaha et~al.}(2001)\citenamefont{Blaha, Schwarz,
  Madsen, Kvasnicka, and Luitz}}]{blaha_wien2k}
\bibinfo{author}{\bibfnamefont{P.}~\bibnamefont{Blaha}},
  \bibinfo{author}{\bibfnamefont{K.}~\bibnamefont{Schwarz}},
  \bibinfo{author}{\bibfnamefont{G.}~\bibnamefont{Madsen}},
  \bibinfo{author}{\bibfnamefont{D.}~\bibnamefont{Kvasnicka}},
  \bibnamefont{and} \bibinfo{author}{\bibfnamefont{J.}~\bibnamefont{Luitz}},
  \emph{\bibinfo{title}{\textsf{Wien2k}, {A}n {A}ugmented {P}lane
  {W}ave+{L}ocal {O}rbitals {P}rogram for {C}alculating {C}rystal
  {P}roperties}} (\bibinfo{publisher}{Tech. Universit\"at Wien, Austria},
  \bibinfo{year}{2001}).

\bibitem[{\citenamefont{Seth et~al.}(2015)\citenamefont{Seth, Hansmann, van
  Roekeghem, Vaugier, and Biermann}}]{Shell-folding}
\bibinfo{author}{\bibfnamefont{P.}~\bibnamefont{Seth}},
  \bibinfo{author}{\bibfnamefont{P.}~\bibnamefont{Hansmann}},
  \bibinfo{author}{\bibfnamefont{A.}~\bibnamefont{van Roekeghem}},
  \bibinfo{author}{\bibfnamefont{L.}~\bibnamefont{Vaugier}}, \bibnamefont{and}
  \bibinfo{author}{\bibfnamefont{S.}~\bibnamefont{Biermann}},
  \bibinfo{journal}{arXiv:1508.07466}  (\bibinfo{year}{2015}).

\bibitem[{\citenamefont{Judd}(1998)}]{judd}
\bibinfo{author}{\bibfnamefont{B.~R.} \bibnamefont{Judd}},
  \emph{\bibinfo{title}{Operator Techniques in Atomic Spectroscopy}}
  (\bibinfo{publisher}{Princeton University Press}, \bibinfo{year}{1998}).

\bibitem[{\citenamefont{Slater}(1960)}]{Slater_book}
\bibinfo{author}{\bibfnamefont{J.~C.} \bibnamefont{Slater}},
  \emph{\bibinfo{title}{Quantum Theory of Atomic Structure}},
  vol.~\bibinfo{volume}{1} (\bibinfo{publisher}{McGraw-Hill, New York},
  \bibinfo{year}{1960}).

\bibitem[{\citenamefont{Sugano et~al.}(1970)\citenamefont{Sugano, Tanabe, and
  Kamimura}}]{Sugano_book}
\bibinfo{author}{\bibfnamefont{S.}~\bibnamefont{Sugano}},
  \bibinfo{author}{\bibfnamefont{Y.}~\bibnamefont{Tanabe}}, \bibnamefont{and}
  \bibinfo{author}{\bibfnamefont{H.}~\bibnamefont{Kamimura}},
  \emph{\bibinfo{title}{Multiplets of transition-metal ions in crystal}},
  vol.~\bibinfo{volume}{1} (\bibinfo{publisher}{Academic Press, New York
  London}, \bibinfo{year}{1970}).

\bibitem[{\citenamefont{Anisimov
  et~al.}(1997{\natexlab{b}})\citenamefont{Anisimov, Aryasetiawan, and
  Lichtenstein}}]{LDA+U-anisimov-1997}
\bibinfo{author}{\bibfnamefont{V.~I.} \bibnamefont{Anisimov}},
  \bibinfo{author}{\bibfnamefont{F.}~\bibnamefont{Aryasetiawan}},
  \bibnamefont{and} \bibinfo{author}{\bibfnamefont{A.~I.}
  \bibnamefont{Lichtenstein}}, \bibinfo{journal}{J. Phys.: Condens. Matter}
  \textbf{\bibinfo{volume}{9}}, \bibinfo{pages}{767}
  (\bibinfo{year}{1997}{\natexlab{b}}).

\bibitem[{\citenamefont{Kutepov et~al.}(2010)\citenamefont{Kutepov, Haule,
  Savrasov, and Kotliar}}]{scGW-kutepov}
\bibinfo{author}{\bibfnamefont{A.}~\bibnamefont{Kutepov}},
  \bibinfo{author}{\bibfnamefont{K.}~\bibnamefont{Haule}},
  \bibinfo{author}{\bibfnamefont{S.~Y.} \bibnamefont{Savrasov}},
  \bibnamefont{and} \bibinfo{author}{\bibfnamefont{G.}~\bibnamefont{Kotliar}},
  \bibinfo{journal}{Phys. Rev. B} \textbf{\bibinfo{volume}{82}},
  \bibinfo{pages}{045105} (\bibinfo{year}{2010}).

\bibitem[{\citenamefont{Antonides et~al.}(1977)\citenamefont{Antonides, Janse,
  and Sawatzky}}]{U-xray-sawatzky-1977}
\bibinfo{author}{\bibfnamefont{E.}~\bibnamefont{Antonides}},
  \bibinfo{author}{\bibfnamefont{E.~C.} \bibnamefont{Janse}}, \bibnamefont{and}
  \bibinfo{author}{\bibfnamefont{G.~A.} \bibnamefont{Sawatzky}},
  \bibinfo{journal}{Phys. Rev. B} \textbf{\bibinfo{volume}{15}},
  \bibinfo{pages}{1669} (\bibinfo{year}{1977}).

\bibitem[{\citenamefont{Sawatzky and Allen}(1984)}]{NiO-sawatzky}
\bibinfo{author}{\bibfnamefont{G.~A.} \bibnamefont{Sawatzky}} \bibnamefont{and}
  \bibinfo{author}{\bibfnamefont{J.~W.} \bibnamefont{Allen}},
  \bibinfo{journal}{Phys. Rev. Lett.} \textbf{\bibinfo{volume}{53}},
  \bibinfo{pages}{2339} (\bibinfo{year}{1984}).

\bibitem[{\citenamefont{van~der Marel et~al.}(1984)\citenamefont{van~der Marel,
  Sawatzky, and Hillebrecht}}]{U-xray-sawatzky-1984}
\bibinfo{author}{\bibfnamefont{D.}~\bibnamefont{van~der Marel}},
  \bibinfo{author}{\bibfnamefont{G.~A.} \bibnamefont{Sawatzky}},
  \bibnamefont{and} \bibinfo{author}{\bibfnamefont{F.~U.}
  \bibnamefont{Hillebrecht}}, \bibinfo{journal}{Phys. Rev. Lett.}
  \textbf{\bibinfo{volume}{53}}, \bibinfo{pages}{206} (\bibinfo{year}{1984}).

\bibitem[{\citenamefont{Lang and Firsov}(1962)}]{Lang_Firsov_transf}
\bibinfo{author}{\bibfnamefont{I.~J.} \bibnamefont{Lang}} \bibnamefont{and}
  \bibinfo{author}{\bibfnamefont{Y.~A.} \bibnamefont{Firsov}},
  \bibinfo{journal}{Sov. Phys. JETP} \textbf{\bibinfo{volume}{16}},
  \bibinfo{pages}{1301} (\bibinfo{year}{1962}).

\bibitem[{\citenamefont{Werner and Millis}(2007)}]{Werner-Holstein}
\bibinfo{author}{\bibfnamefont{P.}~\bibnamefont{Werner}} \bibnamefont{and}
  \bibinfo{author}{\bibfnamefont{A.}~\bibnamefont{Millis}},
  \bibinfo{journal}{Phys. Rev. Lett.} \textbf{\bibinfo{volume}{99}},
  \bibinfo{pages}{146404} (\bibinfo{year}{2007}).

\bibitem[{\citenamefont{Herring}(1966)}]{herring}
\bibinfo{author}{\bibfnamefont{C.}~\bibnamefont{Herring}},
  \emph{\bibinfo{title}{Magnetism: Exchange interactions among itinerant
  electrons}}, vol.~\bibinfo{volume}{4} (\bibinfo{publisher}{Academic Press},
  \bibinfo{year}{1966}).

\bibitem[{\citenamefont{Yamasaki et~al.}(2010)\citenamefont{Yamasaki, Matsui,
  Imada, Takase, Azuma, Muro, Kato, Higashiya, Sekiyama, Suga
  et~al.}}]{Yamasaki-FeSe}
\bibinfo{author}{\bibfnamefont{A.}~\bibnamefont{Yamasaki}},
  \bibinfo{author}{\bibfnamefont{Y.}~\bibnamefont{Matsui}},
  \bibinfo{author}{\bibfnamefont{S.}~\bibnamefont{Imada}},
  \bibinfo{author}{\bibfnamefont{K.}~\bibnamefont{Takase}},
  \bibinfo{author}{\bibfnamefont{H.}~\bibnamefont{Azuma}},
  \bibinfo{author}{\bibfnamefont{T.}~\bibnamefont{Muro}},
  \bibinfo{author}{\bibfnamefont{Y.}~\bibnamefont{Kato}},
  \bibinfo{author}{\bibfnamefont{A.}~\bibnamefont{Higashiya}},
  \bibinfo{author}{\bibfnamefont{A.}~\bibnamefont{Sekiyama}},
  \bibinfo{author}{\bibfnamefont{S.}~\bibnamefont{Suga}}, \bibnamefont{et~al.},
  \bibinfo{journal}{Phys. Rev. B} \textbf{\bibinfo{volume}{82}},
  \bibinfo{pages}{184511} (\bibinfo{year}{2010}).

\bibitem[{\citenamefont{Brouet et~al.}(2010)\citenamefont{Brouet,
  Rullier-Albenque, Marsi, Mansart, Aichhorn, Biermann, Faure, Perfetti,
  Taleb-Ibrahimi, Le~F\`evre et~al.}}]{BaFeAs-brouet}
\bibinfo{author}{\bibfnamefont{V.}~\bibnamefont{Brouet}},
  \bibinfo{author}{\bibfnamefont{F.}~\bibnamefont{Rullier-Albenque}},
  \bibinfo{author}{\bibfnamefont{M.}~\bibnamefont{Marsi}},
  \bibinfo{author}{\bibfnamefont{B.}~\bibnamefont{Mansart}},
  \bibinfo{author}{\bibfnamefont{M.}~\bibnamefont{Aichhorn}},
  \bibinfo{author}{\bibfnamefont{S.}~\bibnamefont{Biermann}},
  \bibinfo{author}{\bibfnamefont{J.}~\bibnamefont{Faure}},
  \bibinfo{author}{\bibfnamefont{L.}~\bibnamefont{Perfetti}},
  \bibinfo{author}{\bibfnamefont{A.}~\bibnamefont{Taleb-Ibrahimi}},
  \bibinfo{author}{\bibfnamefont{P.}~\bibnamefont{Le~F\`evre}},
  \bibnamefont{et~al.}, \bibinfo{journal}{Phys. Rev. Lett.}
  \textbf{\bibinfo{volume}{105}}, \bibinfo{pages}{087001}
  (\bibinfo{year}{2010}).

\bibitem[{\citenamefont{Vildosola et~al.}(2008)\citenamefont{Vildosola,
  Pourovskii, Arita, Biermann, and Georges}}]{LaOFeAs-veronica}
\bibinfo{author}{\bibfnamefont{V.}~\bibnamefont{Vildosola}},
  \bibinfo{author}{\bibfnamefont{L.}~\bibnamefont{Pourovskii}},
  \bibinfo{author}{\bibfnamefont{R.}~\bibnamefont{Arita}},
  \bibinfo{author}{\bibfnamefont{S.}~\bibnamefont{Biermann}}, \bibnamefont{and}
  \bibinfo{author}{\bibfnamefont{A.}~\bibnamefont{Georges}},
  \bibinfo{journal}{Phys. Rev. B} \textbf{\bibinfo{volume}{78}},
  \bibinfo{pages}{064518} (\bibinfo{year}{2008}).

\bibitem[{\citenamefont{Pines}(1964)}]{pines-plasmons}
\bibinfo{author}{\bibfnamefont{D.}~\bibnamefont{Pines}},
  \emph{\bibinfo{title}{Elementary excitations in solids : lectures on phonons,
  electrons, and plasmons}} (\bibinfo{publisher}{W.A. Benjamin},
  \bibinfo{year}{1964}).

\bibitem[{\citenamefont{Nomura et~al.}(2012)\citenamefont{Nomura, Kaltak,
  Nakamura, Taranto, Sakai, Toschi, Arita, Held, Kresse, and
  Imada}}]{Nomura_local_U}
\bibinfo{author}{\bibfnamefont{Y.}~\bibnamefont{Nomura}},
  \bibinfo{author}{\bibfnamefont{M.}~\bibnamefont{Kaltak}},
  \bibinfo{author}{\bibfnamefont{K.}~\bibnamefont{Nakamura}},
  \bibinfo{author}{\bibfnamefont{C.}~\bibnamefont{Taranto}},
  \bibinfo{author}{\bibfnamefont{S.}~\bibnamefont{Sakai}},
  \bibinfo{author}{\bibfnamefont{A.}~\bibnamefont{Toschi}},
  \bibinfo{author}{\bibfnamefont{R.}~\bibnamefont{Arita}},
  \bibinfo{author}{\bibfnamefont{K.}~\bibnamefont{Held}},
  \bibinfo{author}{\bibfnamefont{G.}~\bibnamefont{Kresse}}, \bibnamefont{and}
  \bibinfo{author}{\bibfnamefont{M.}~\bibnamefont{Imada}},
  \bibinfo{journal}{Phys. Rev. B} \textbf{\bibinfo{volume}{86}},
  \bibinfo{pages}{085117} (\bibinfo{year}{2012}).

\bibitem[{\citenamefont{Steiner et~al.}(2015)\citenamefont{Steiner, Nomura, and
  Werner}}]{Steiner_double_expansion}
\bibinfo{author}{\bibfnamefont{K.}~\bibnamefont{Steiner}},
  \bibinfo{author}{\bibfnamefont{Y.}~\bibnamefont{Nomura}}, \bibnamefont{and}
  \bibinfo{author}{\bibfnamefont{P.}~\bibnamefont{Werner}},
  \bibinfo{journal}{Phys. Rev. B} \textbf{\bibinfo{volume}{92}},
  \bibinfo{pages}{115123} (\bibinfo{year}{2015}).

\end{thebibliography}

\end{document}